\documentclass[10pt,letterpaper]{article}

\usepackage{dsfont}
\usepackage{epstopdf}
\usepackage{acronym}
\usepackage{graphicx}
\usepackage{float}
\usepackage{subfig}
\usepackage[T1]{fontenc}
\usepackage[utf8]{inputenc}
\usepackage{authblk}
\usepackage{amsmath}
\usepackage{amsfonts}
\usepackage{color}
\usepackage{amssymb}
\usepackage{algorithm}
\usepackage[margin=0.7in]{geometry}

\begin{document}

\title{No equations, no parameters, no variables: \\
data, and the reconstruction of normal forms \\
by learning informed observation geometries}

\author[a]{Or Yair}
\author[a,1]{Ronen Talmon}
\author[b,1]{Ronald R. Coifman}
\author[c,1]{Ioannis G. Kevrekidis}

\affil[a]{Viterbi Faculty of Electrical Engineering, Technion - Israel Institute of Technology, Haifa 32000, Israel}
\affil[b]{Department of Mathematics, Yale University, New Haven, CT 06511, USA}
\affil[c]{Department of Chemical and Biological Engineering and PACM, Princeton University, Princeton, NJ 08544, USA}

\maketitle

\begin{abstract}
The discovery of physical laws consistent with empirical observations
lies at the heart of (applied) science and engineering.
These laws typically take the form of nonlinear differential equations depending on parameters;
dynamical systems theory provides, through the appropriate normal forms, an ``intrinsic'', prototypical
characterization of the types of dynamical regimes accessible to a given model.
Using an implementation of data-informed geometry learning we directly reconstruct the relevant ``normal forms'':
a quantitative mapping from empirical observations to prototypical realizations of the underlying dynamics.
Interestingly, the state variables and the parameters of these realizations are inferred from the empirical
observations; without prior knowledge or understanding,
they parametrize the dynamics {\em intrinsically}, without explicit reference to fundamental physical quantities.

\end{abstract}

\section*{Introduction}

\begin{figure}[t]
\centering \includegraphics[height=5.5cm]{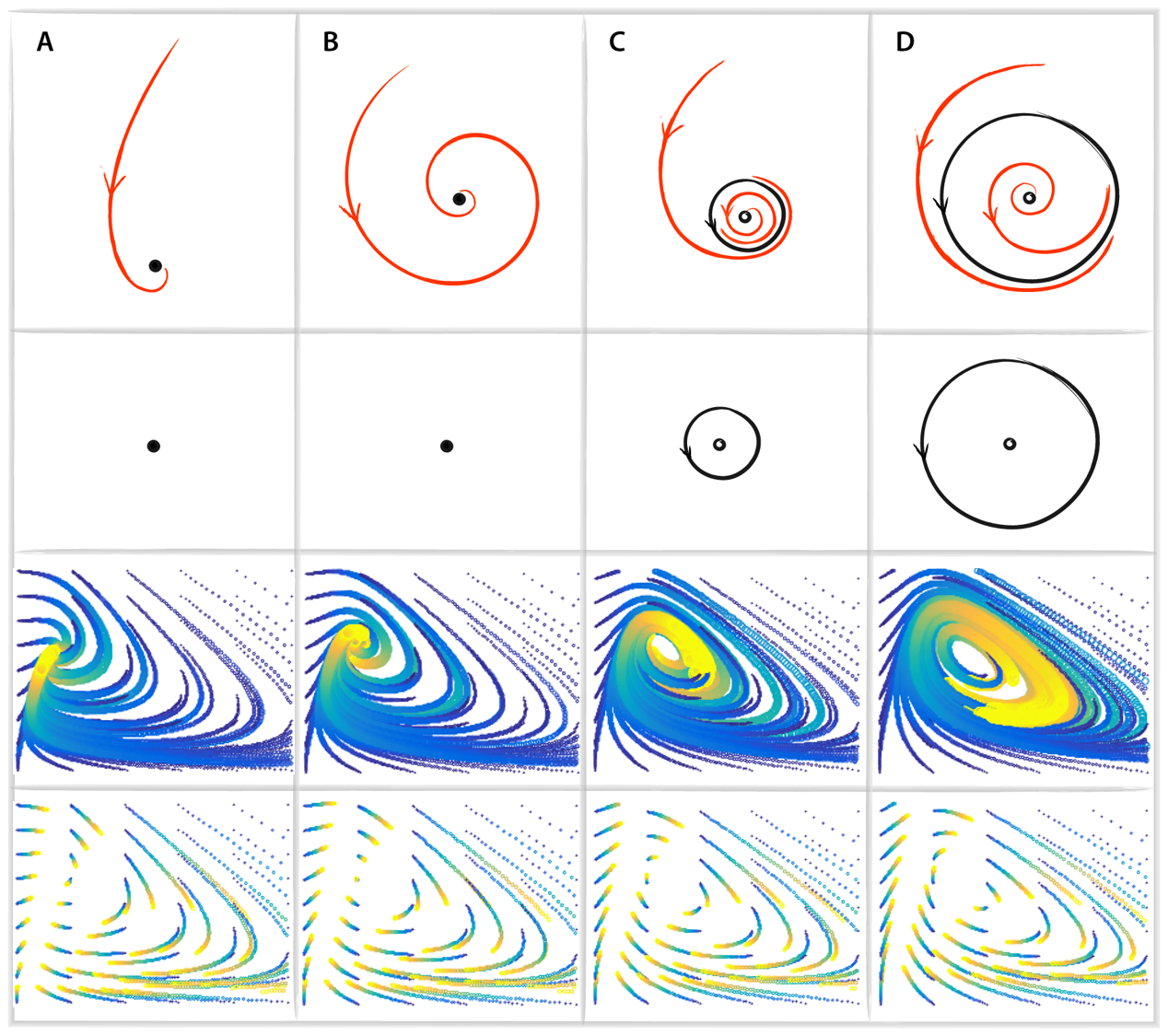}
\protect\caption{Inferring a Hopf bifurcation from phase portrait representations (see also text).}
\label{fig:Hopf_cards}
\end{figure}

Consider the four sketches displayed in the first row of Figure \ref{fig:Hopf_cards}.
Solely by observation, one could recognize that each image is a phase-portrait in a sequence involving a Hopf bifurcation.
Taking into account the four images together, one might even identify the trend -- that this sequence should be associated with a supercritical Hopf bifurcation, where a stable limit cycle is ``born'' somewhere between Panel B and Panel C as the steady state loses stability at some critical parameter value.
It is ``obvious'' from the transients that the steady state becomes less attracting between Panel A and Panel B; the amplitude of the limit cycle gradually grows away from the critical parameter value, and its period should be related to the rate of spiraling of the transients around the steady states.

If we now erased the labels of the panels (A--D), and shuffled the four cards, one could easily figure out how to put them back in the ``right'' order (in
effect understanding that this is ``naturally'' a one-parameter family of images).
By recalling the normal form of a Hopf bifurcation, one could attempt to fit some coefficients, find the period and its rate of change with the parameter, and, arguably, even estimate where to quantitatively pin Panel C in a parametric interval whose edges are Panel B and Panel D.

This caricature demonstrates our ability to recognize dynamic patterns, to focus on the salient features, and to make predictions based on data.
In contrast, the sketches in the second row of Figure \ref{fig:Hopf_cards} do not give enough information to discriminate between Panel A and Panel B, while the sketches in the third and
fourth rows require some additional ``mental image processing'' in order to figure out when a limit cycle exists and which limit cycle is bigger, revealing the parameter space ordering.

The purpose of this paper is to present a particular implementation of a set of algorithms that allow a systematic realization of all these steps -- inferring ``normal forms'' from data, using just the data-mining counterpart of the above discussion: similarity between nearby observations/measurements.
In the third row of Figure \ref{fig:Hopf_cards} we plot several simulated trajectories from a uniform grid of initial conditions in state space. Observing these data, one can easily infer the dynamical regime and deduce the underlying one-parameter family, in a similar manner to the exercise above.
Our methodology aspires to accomplish much more than what one can deduce based on such ``easy'' observations.
For example, in the fourth row of Figure \ref{fig:Hopf_cards}, we present the same type of data as in the third row, but depicting far shorter trajectories. Now the four images look much more similar - without prior knowledge or additional post-processing, it would be challenging to correctly infer the dynamical regime and the one-parameter family from observations.
Our methodology will be successful even with these short trajectory data.

\begin{figure}[t]
\centering \includegraphics[height=4cm]{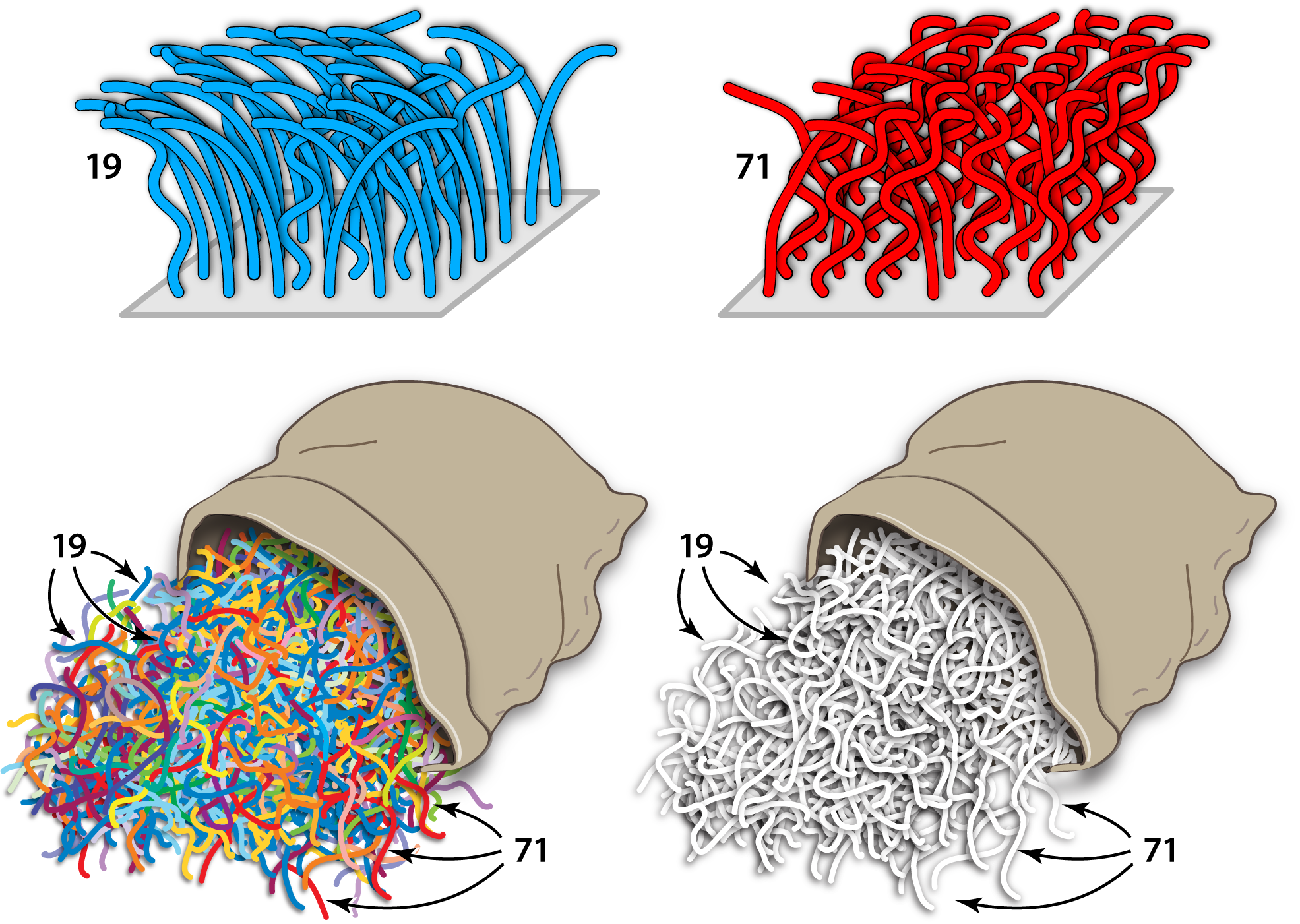}
\caption{We observe an ensemble of short trajectories in what we call {\em a trial} from a fixed, yet unknown, dynamical regime. For example, one trial is labeled $19$ and another trial is labeled $71$. We store the short trajectories (``bag'' them), and we only keep the label of the trial. Our goal is to empirically derive the dynamical regime information associated with each trajectory (labelled by the color in this case), by (a) deducing the state variables and the associated phase portraits; and then (b) organizing the phase-portraits of all the trials; so as to (c) derive (a tabulated form of) the evolution equation governing the dynamics.}
\label{fig:carpet}
\end{figure}
In the language of systems theory we consider multiple measurements, at different parameter settings -- what we call different {\em trials}, from an unknown, nonlinear, parametrically-dependent dynamical system.
The problem setting comprises a large ensemble of short time series, indexed by the label of the trial as well as the label of the measurement channel;
each time series is parametrized by time.
However, the knowledge of how many and what parameters the system has, and the actual settings at which the trials are performed, is hidden. Furthermore, we will not know how many and what state variables the system has, nor what  functions of the state variables we are measuring. We will only know what each channel recorded, at each trial, as a function of time (for a short time).
A caricature of this problem setup is depicted in Figure \ref{fig:carpet}.
Using the pairwise similarity between individual pairs of this large ensemble of short time series as our only tool, we empirically build a ``normal form'' of the system: we identify a set of relevant parameters, a set of relevant state variables, and help reveal the nature and relation of the associated
phase portraits.

Recovering the underlying structure of nonlinear dynamical systems from data (``system identification'') has attracted significant research efforts over many years, and several ingenious techniques have been proposed to address different aspects of this problem.
These include methods to find nonlinear differential equations \cite{bongard2007automated,schmidt2009distilling},
to discover governing equations from time-series \cite{crutchfield1987equations,brunton2016discovering},
equation-free modeling approaches \cite{kevrekidis2003equation}, and methods for empirical dynamic modeling \cite{sugihara2012detecting}, just to name a few.
In this paper, we present a technique with roots in manifold learning \cite{Tenenbaum:2000,Roweis:2000,Donoho2003,Belkin2003}, which involves tensor-geometry learning \cite{Gavish2012,Ankenman2014,mishne2015hierarchical} as well as metric learning and approximation \cite{Leeb2015}.

\section*{Problem Formulation and a Toy Example}

Our purpose is to devise a data-driven framework for the organization
of time-dependent observations of dynamical systems depending on parameters.
In our setting these time-dependent measurements are the result of
a number of experiments that we will call {\em trials}; during each
trial, the (unknown) parameter values remain constant.
In this {\em black box} setting, the dynamical system is unknown, nonlinear
and {\em autonomous}, and is given by
\begin{align}
\frac{d\boldsymbol{x}}{dt} & =f(\boldsymbol{x};\boldsymbol{p})\\
\boldsymbol{y} & =h(\boldsymbol{x})
\end{align}
We do not have access to its state $\boldsymbol{x}$ nor to its parameter values $\boldsymbol{p}$; we also do not know
the evolution law $f$, nor the measurement function $h$. We only
have measurements $\boldsymbol{y}$ labelled by time $t$.
The black box is endowed with ``knobs'' that, in an
unknown way, change the values of the parameters $\boldsymbol{p}$; so in every
trial, for a new, but unknown, set of parameter values $\boldsymbol{p}$, we can
observe $\boldsymbol{y}$ coming out of the box without knowing $\boldsymbol{x}$ or $f$ or even $h$.
We want to characterize the system dynamics by systematically
organizing our observations (collected over several trials) of its
outputs.

More specifically, we want to (a) organize the observations by finding
a set of \em{state variables} and a set of
\em{system parameters} that jointly preserve the essential features of the dynamics;
and then (b) find the corresponding {\em intrinsic}
geometry of this combined variable-parameter space, thus building a sort of
normal form for the problem.
Small changes in this {\em jointly intrinsic space}
will correspond to small changes in dynamic behavior (i.e. to robustness).
Having discovered a useful
``joint geometry'' we can then inspect its individual
constituents. Inspecting, for example, the geometry of the discovered
parameter space, will help identify regimes of different qualitative
behavior. This might be different dynamic behavior, like hysteresis, or oscillations, separated by bifurcations; alternatively,
we might observe transitions between different sizes of the minimal realizations:
regimes where the number of minimal variables/parameters necessary in the realization
changes.

We can also inspect the identified state variable geometry, which
will help us organize the temporal measurements in coherent phase
portraits. In addition, if there exist regimes where the system becomes
\emph{singularly perturbed}, we expect we will be able to realize that the requisite minimal phase
portrait dimension changes (reduces), and that the reduction in the
number of state variables is linked with the reduction in the number
of intrinsic parameters.

As an illustrative example, consider the following dynamical system, arising in the unfolding of the Bogdanov-Takens singularity \cite{guckenheimer2013nonlinear}:
\begin{align}\label{eq:BT}
\frac{dx_1}{dt} & =x_2 \nonumber \\
\frac{dx_2}{dt} & =\beta_{1}+\beta_{2}x_1+x_1^{2}-x_1x_2.
\end{align}

This set of differential equations defines a dynamical system with two parameters
$\boldsymbol{p}=(\beta_{1},\beta_{2})$, two state variables $\boldsymbol{x}=(x_{1},x_{2})$, and two observables $\boldsymbol{y}=(y_{1},y_{2})$;
at first we choose the observable to be the state variables themselves, i.e., $(y_{1},y_{2})=(x_{1},x_{2})$ with $h(\boldsymbol{x})$ being the identity function.
It is known that the parameter space of this system $(\beta_{1},\beta_{2})$
can be divided into $4$ different regimes separated by one-parameter bifurcation curves \cite{guckenheimer2013nonlinear}.
Figure~\ref{fig:BT_params}~(left) shows this ``ground truth'' bifurcation diagram for our simulated $2$D
grid of parameter values. Each point $\boldsymbol{p}=(\beta_{1},\beta_{2})$ on the grid is colored according to its respective dynamical regime.

Our goal in this case would be to discover an accurate bifurcation
map of the system in a data-driven manner purely from observations.
These observations consist of several samples, where each sample is a single trajectory $\boldsymbol{y}(t)$ of the system initialized with unknown (possibly different) parameter values and initial values.
In addition, we would like to deduce from these large number of realizations of trajectories $\boldsymbol{y}(t)$ arbitrarily and differently initialized that the system depends on only two parameters and can be realized with only two state variables; and to reconstruct the bifurcation diagram with its
phase portraits.

\section*{Geometry learning of dynamics from observations}

Consider data arising from an autonomous dynamical system; we view the observations
as entries in a three-dimensional tensor. One axis of the tensor corresponds
to variations in the problem parameters, one to variations in the
problem variables, and
the third axis corresponds to time evolution along trajectories.

Formally, let $\mathcal{P}$ denote an ensemble of $N_{p}$ sets of the
$d_{p}$ system parameters. Let $\mathcal{V}$ be
a ensemble of $N_{v}$ sets of initial condition values  of the $d_{v}$ state variables.
For each $\boldsymbol{p}\in\mathcal{P}$
and $\boldsymbol{v}\in\mathcal{V}$, we observe a trajectory
$Y(\boldsymbol{v},\boldsymbol{p},t)$ of length $N_{t}$ in $\mathbb{R}^{d_v}$ of the system
variables, where $t=1,\ldots,N_t$ denotes the time sample.
In summary, $\boldsymbol{p}$ is a label of the particular differential equations of the dynamical system, $\boldsymbol{v}$ is a label of the observations trajectory, and $t$ is the time label.

Let $\mathbf{Y}$ denote the
entire $3$D tensor of observations of dimension $N_{p}\times N_{v}\times N_{t}$
consisting of all the data at hand.
With respect to the black box setting described in the introduction, we emphasize that the identity of the parameters and variables is hidden; we only have trajectories of observations corresponding to various trials with possibly different hidden parameter values and with different hidden initial input coordinates.

\begin{figure}[t]
\centering
\includegraphics[width=0.9\columnwidth]{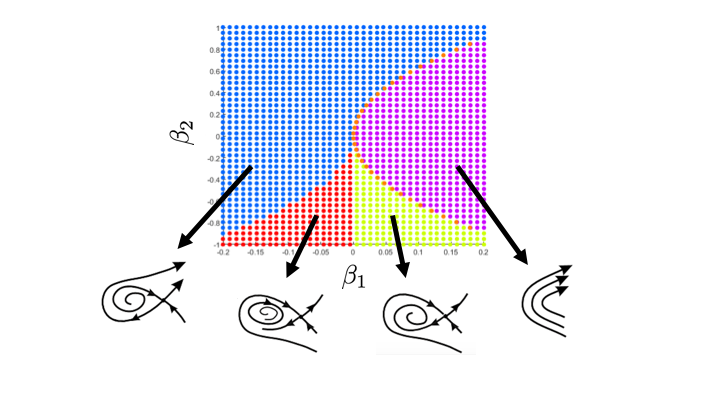}
\includegraphics[height=3cm]{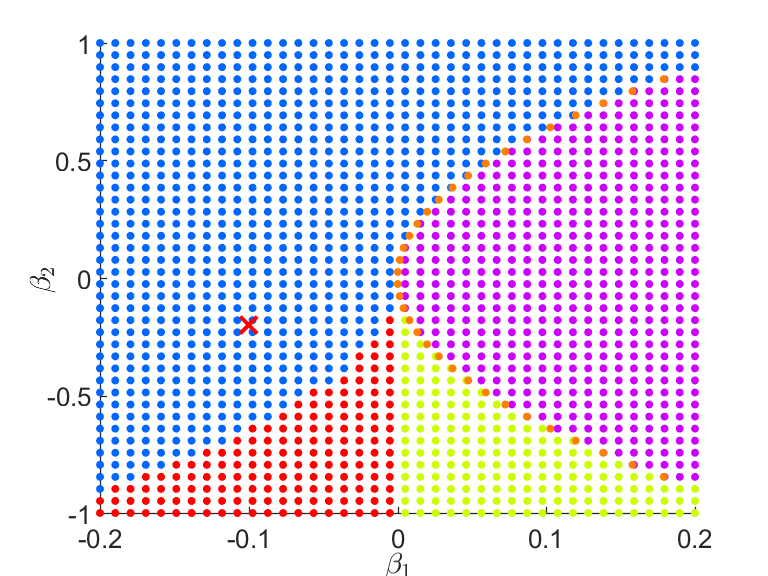}
\includegraphics[height=2.5cm]{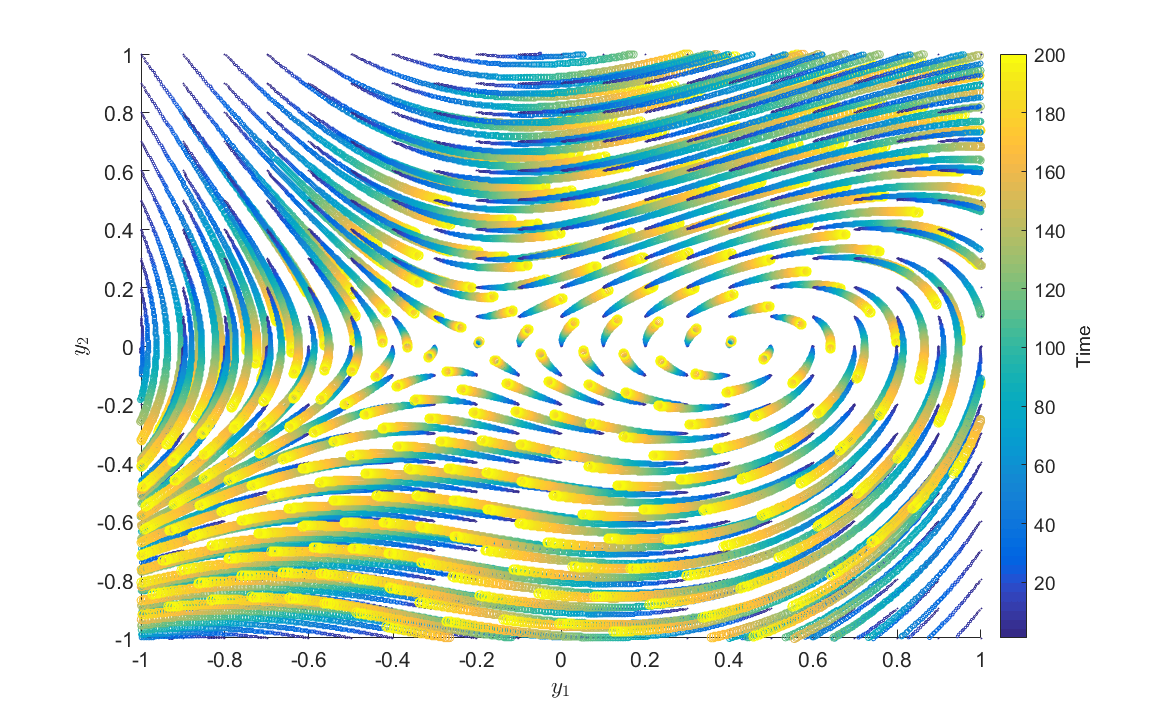}

\caption{(up) The Bogdanov-Takens bifurcation maps with insets illustrating the typical phase-portraits in each dynamical regime.
	     (left) The Bogdanov-Takens bifurcation map. (right) An example of the phase-portrait of the simulated trajectories of the Bogdanov-Takens system corresponding to the parameter set $(\beta_{1}, \beta_{2})=(-0.1,-0.2)$, marked by red 'x' on the left. The color and the width of the point corresponds to time.}
\label{fig:BT_params}
\end{figure}

To make the problem definition concrete we describe the setting of a specific example.
Recall the Bogdanov-Takens dynamical system of two variables and two parameters, introduced in \eqref{eq:BT}.
We generate a set $\mathcal{P}$ of $N_p = 400$ different parameter values $\boldsymbol{p}=(\beta_{1},\beta_{2})$
from a regular fixed $2$D grid, where $\beta_{1}\in\left[-0.2,0.2\right]$ and $\beta_{2}\in\left[-1,1\right]$, and additional $10$ parameter values located exactly on the bifurcation.
Similarly, we generate a set $\mathcal{V}$
of $N_v = 441$ different initial conditions $\boldsymbol{v}=(y_{1}(0),y_{2}(0))$ from
a fixed $2$D grid in $\left[-1,1\right]^{2}$.
For each $\boldsymbol{p}\in\mathcal{P}$ and $\boldsymbol{v}\in\mathcal{V}$,
we observe a trajectory of the system for $N_{t}=200$ time steps,
where the interval between two adjacent time samples is
$\Delta t=0.004\,\mbox{[sec]}$ and collect all the trajectories into a single $3$D tensor $\mathbf{Y}$.
In this example, $N_p = 410, N_v = 441$ and $N_t = 200$ so overall we have
$\mathbf{Y} \in \mathbb{R}^{410 \times 441 \times 200}$.
For illustration purposes, Figure \ref{fig:BT_params}~(right) depicts
the phase portrait of all the simulated 2D trajectories from all the
initial points for {\em a single} particular fixed value of $\boldsymbol{p}$,
$\beta_{1}=-0.1$ and $\beta_{2}=-0.2$ (marked by a red `{\color{red}$\times$}' in Figure \ref{fig:BT_params}~(left)).
We note that the trajectories (as illustrated in Figure \ref{fig:BT_params}) are long enough to partially overlap in phase space.
Such an overlap induces the coupling between the time and variables axes, which is captured and exploited by our analysis.
We wish to find a reliable representation of the hidden parameters, of the hidden variables, and of the time axis.

Define $\boldsymbol{y}_{p}=\left\{ Y(\boldsymbol{v},\boldsymbol{p},t)|\forall\boldsymbol{v},\forall t\right\}$ for each of the $N_{p}$ vectors of hidden parameter values $\boldsymbol{p}$ in $\mathcal{P}$, namely, a data sample consisting of all the trajectories from a single trial.
For simplicity of notation, we will use subscripts to denote both the appropriate axis and a specific set of entries values on the axis.
We refer to $\{\boldsymbol{y}_{p}\}, \boldsymbol{p} \in \mathcal{P}$ as the data samples from the parameters axis viewpoint.
In the Bogdanov-Takens example, Figure \ref{fig:BT_params} depicts
$\boldsymbol{y}_p$ for $\boldsymbol{p}=(\beta_{1},  \beta_{2}) = (-0.1,-0.2)$.

Similarly, let $\boldsymbol{y}_{v}$ and $\boldsymbol{y}_{t}$ be the samples from the viewpoints of the variables axis and the time axis, respectively, which are defined by
\begin{align*}
\boldsymbol{y}_{v} & =\left\{ Y(\boldsymbol{v},\boldsymbol{p},t)|\forall\boldsymbol{p},\forall t\right\}, \quad \boldsymbol{v} \in \mathcal{V} \\
\boldsymbol{y}_{t} & =\left\{ Y(\boldsymbol{v},\boldsymbol{p},t)|\forall\boldsymbol{v},\forall\boldsymbol{p}\right\}, \quad t=1,\ldots,N_t.
\end{align*}
One way to accomplish our goal is to process the data {\em three successive times}, each time from a different viewpoint.

Here, we use a data-driven parametrization approach based on a kernel.
From the trials (effectively, parameters) axis point of view, a typical kernel is defined by
\begin{equation}
	k(\boldsymbol{y}_{p_1}, \boldsymbol{y}_{p_2}) = e^{-\|\boldsymbol{y}_{p_1}-\boldsymbol{y}_{p_2}\|^{2}/\epsilon}, \forall \boldsymbol{p}_1, \boldsymbol{p}_2 \in \mathcal{P}
\label{eq:affinity_matrix}
\end{equation}
based on distances between any pair of samples, where the Gaussian function induces a sense of locality relative to the kernel scale $\epsilon$. To aggregate the pairwise affinities comprising the kernel into a global parametrization,
traditionally, the eigenvalue decomposition (EVD) is applied to the kernel,
and the eigenvectors are used as the desired parametrization. For more details, see Appendix \ref{Sec:Diffuion_Maps}.

From three {\em separate} diffusion maps applications to the sets $\{\boldsymbol{y}_{p}\}$, $\{\boldsymbol{y}_{v}\}$, and $\{\boldsymbol{y}_{t}\}$, we can obtain three mappings as in \eqref{eq:dmaps}, denoting the associated eigenvectors by $\{ \boldsymbol{\psi}_\ell^{\mathcal{P}} \}$, $\{ \boldsymbol{\psi}_\ell^{\mathcal{V}} \}$, and $\{ \boldsymbol{\psi}_\ell^{\mathcal{T}} \}$, respectively.

However, such mappings do not take into account the strong correlations and co-dependencies between the parameter values and the dynamics of the variables which arise in typical dynamical systems.
For example, in the Bogdanov-Takens system, the dynamical regime changes significantly depending on the values of the parameters.

To incorporate such co-dependencies, we propose to define and build from observations an {\em informed metric} between samples in the different axes.
In the introduction of the affinity matrix in \eqref{eq:affinity_matrix},
we deliberately did not specify the norm used to compare between two
samples. Common practice is to use the Euclidean norm. However, as
pointed out by Lafon \cite{LafonThesis}, {\em anisotropic} diffusion
maps can be computed by using different norms. This issue has been
extensively studied recently, and several norms and metrics have been
developed for this purpose, e.g., \cite{Singer2008,giannakis2015dynamics,dsilva2016data}.

Here, following \cite{Coifman2011,Gavish2012,Ankenman2014,mishne2015hierarchical}, we propose a particular construction
of an {\em informed metric} that relies on the geometry of the
{\em coordinates} of the samples; the metric and the (induced) geometry evolve together, as will be described below.
For simplicity, the exposition begins by focusing on the analysis
of the data from the perspective of the parameters axis; the generalization
to the other two axes is analogous.

The essence of our analysis is the definition of a meaningful notion
of distance between the samples. Specifically, we build
{\em an informed distance metric} $\|\boldsymbol{y}_{p_1}-\boldsymbol{y}_{p_2}\|_{\mathcal{P}}$,
where the subscript of the norm $\mathcal{P}$ indicates that it is an informed
norm between the samples - informed {\em from the parameters viewpoint}.
The construction of the metric is implemented in an iterative manner.
The idea is that in each iteration, the co-dependencies between the axes are gradually revealed from observations and in turn are used to build a refined informed metric.
The full details of the construction procedure are presented in Appendix \ref{appx:informed_metric}.
Here we only describe the first iteration when applied to data arising from the Bogdanov-Takens system.

In the first iteration, the construction of the informed metric $\| \cdot \|_{\mathcal{P}}$ defined on the parameters axis uses as an initial input two {\em non-informed} metrics, defined on the variables axis and on the time axis, whose roles will be made clear in the sequel. Possible choices for such metrics are the Euclidean metric or a metric derived from the cosine similarity.

The construction itself is implemented by decomposing
the metric into the following general form
\begin{equation}
\|\boldsymbol{y}_{p_1}-\boldsymbol{y}_{p_2}\|_{\mathcal{P}}=\|\boldsymbol{y}_{p_1}-\boldsymbol{y}_{p_2}\|_{1}+\gamma\| \mathcal{F}_{\mathcal{P}}(\boldsymbol{y}_{p_1})-\mathcal{F}_{\mathcal{P}}(\boldsymbol{y}_{p_2})\|_{1}
\end{equation}
where $\|\cdot\|_{1}$ is the $\ell_{1}$ norm, $\gamma>0$ is a positive
weighting factor, and $\mathcal{F}_{\mathcal{P}}:\mathbb{R}^{N_{v}\times N_{t}}\rightarrow\mathbb{R}^{D_{p}}$ is some feature function (to be discussed).
We note that the particular choice of the $\ell _1$ norm is explained in detail in \cite{Leeb2015} and will be reviewed in Appendix \ref{appx:informed_metric}, yet other norms can be used depending on the application at hand.
The function $\mathcal{F}_{\mathcal{P}}$ is therefore viewed as a generalized
transform of the samples, and the problem of finding a meaningful
metric is transformed to the problem of finding an appropriate transformation
$\mathcal{F}_{\mathcal{P}}$. In this work, we present a transform {\em that appends coordinates to the samples},  such that the $\ell _1$ norm is equivalent to a generalized \ac{EMD} \cite{Leeb2015}.
Since the \ac{EMD} (in contrast, for example, to the Euclidean distance) takes into account the ``ground distance'' \cite{rubner2000earth}, it is stable under small deformations of the data. This property is important in the context of dynamical systems, as it allows for the comparison of similar trajectories {\em even when partially overlapping}.

Traditional (two-dimensional) transforms are typically implemented using a set of basis functions $g_{\ell,\ell'}$, and the transform is generally given by a collection of the linear projections of the data on that set of basis functions
\begin{equation}
\mathcal{F}_{\mathcal{P}}(\boldsymbol{y}_{p})=\left\{ \langle g_{\ell,\ell'},\boldsymbol{y}_{p}\rangle |\forall \ell \right\}.	
\end{equation}
In particular for the parameters axis, the basis functions $g_{\ell,\ell'}$ are defined on $\mathcal{V} \times \{ 1,\ldots, N_t \}$, and the inner product is defined by
\begin{equation*}
	\langle g_{\ell,\ell'},\boldsymbol{y}_{p}\rangle = \sum _{\boldsymbol{v} \in \mathcal{V}} \sum _{t=1,\ldots, N_t} g_{\ell,\ell'} (\boldsymbol{v}, t) Y(\boldsymbol{p},\boldsymbol{v},t).
\end{equation*}
Such classical transforms include the Fourier Transform, the Wavelet transform, etc. However, these transforms suffer from one or more of the following limitations: their basis functions $g_\ell$ are fixed and not data-adaptive, they are linear, and they are local.

To circumvent these limitations, following \cite{Ankenman2014,mishne2015hierarchical}, we propose
a transform based on data-driven partition trees.
By using the initial, non-informed metrics on the variables and on the parameters, multilevel partitions of both the variables axis and time axis are computed. In turn, these trees are used to define an over-complete set of basis functions; a basis function is defined for each folder $I_\ell$ in the variables tree and for each folder $J_{\ell'}$ in the time axis tree, as the indicator function for the samples in these folders, i.e.,
\begin{equation*}
	g_{\ell,\ell'}(\boldsymbol{v},t) = \left\{ \begin{array}{cc} 1 & \quad \boldsymbol{v} \in I_\ell, t \in J_{\ell'} \\ 0 & \quad \textmd{otherwise.} \end{array} \right.
\end{equation*}
Once we have the basis functions, the transform can be formulated, and based on that, the \ac{EMD} can be defined.
For details, see Appendix \ref{appx:informed_metric}.

Note that the partition trees provide a specific set of basis functions; yet any basis function set could be used for this purpose.
A particular alternative implementation is via the eigenvectors of diffusion maps \cite{Coifman2006}. Broadly, the informed metric is implemented by a generalized transform that appends coordinates to the original samples, such that a non-informed norm/distance between the resulting, {\em augmented/transformed samples} retains particularly desired properties.
Since the outer products of the eigenvectors obtained by diffusion maps $\{ \boldsymbol{\psi}_\ell^{\mathcal{V}} \otimes \boldsymbol{\psi}_{\ell'}^{\mathcal{T}} \}$ are defined on $\mathcal{V} \times \{1,\ldots,N_t\}$, they can be used as a set of basis functions of $\{\boldsymbol{y}_p\}$, and additional coordinates can be appended to the samples by projections on these functions. This alternative implementation is being currently explored.

Several remarks are due at this point.
First, the recursive procedure described above repeats in iterative manner, where in each iteration, three informed metrics are constructed one by one, based on the metrics {\em from the preceding iteration}. As the iterations progress, the metrics are gradually refined, and the dependency on the initialization is reduced.
The outline of the recursive algorithm is presented in Algorithm \ref{algo1}.
Similar iterative approaches for decomposing $3$-tensors are based on iterative singular value decompositions obtained by isolating an axis \cite{comon2009tensor}.
Second, the chcracterization of the stopping criterion and the convergence of this iterative procedure should be possible through the combination of asymptotic
analysis of the informed kernels \cite{Coifman2006} with the alternating $\ell_1$ minimization in \cite{chi2016convex,anandkumar2014tensor},
Currently, we typically apply a few iterations (in this paper, up to two); this has empirically led to good performance.
Third, the exposition here focuses on transforms based on {\em linear} projections on basis functions.
We note that nonlinear embedding constructed from the basis functions themselves, e.g., diffusion maps \cite{Coifman2006}, can also be used as the additional coordinates appended to the samples. This point, as well as additional technical details, will be further discussed in Appendix \ref{appx:informed_metric}.

\section*{Examples}
The code is available at \cite{code}.

\subsection*{Bogdanov-Takens Bifurcation Mapping}

\begin{figure*}[t]

\subfloat[]{\includegraphics[height=3.3cm]{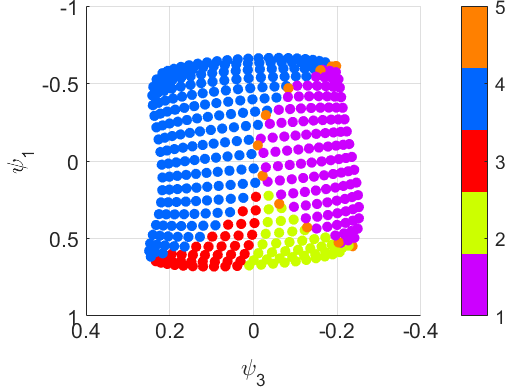}}
\subfloat[]{\includegraphics[height=3.3cm]{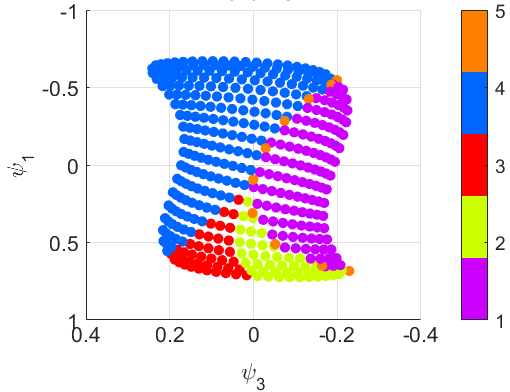}}
\subfloat[]{\includegraphics[height=3.3cm]{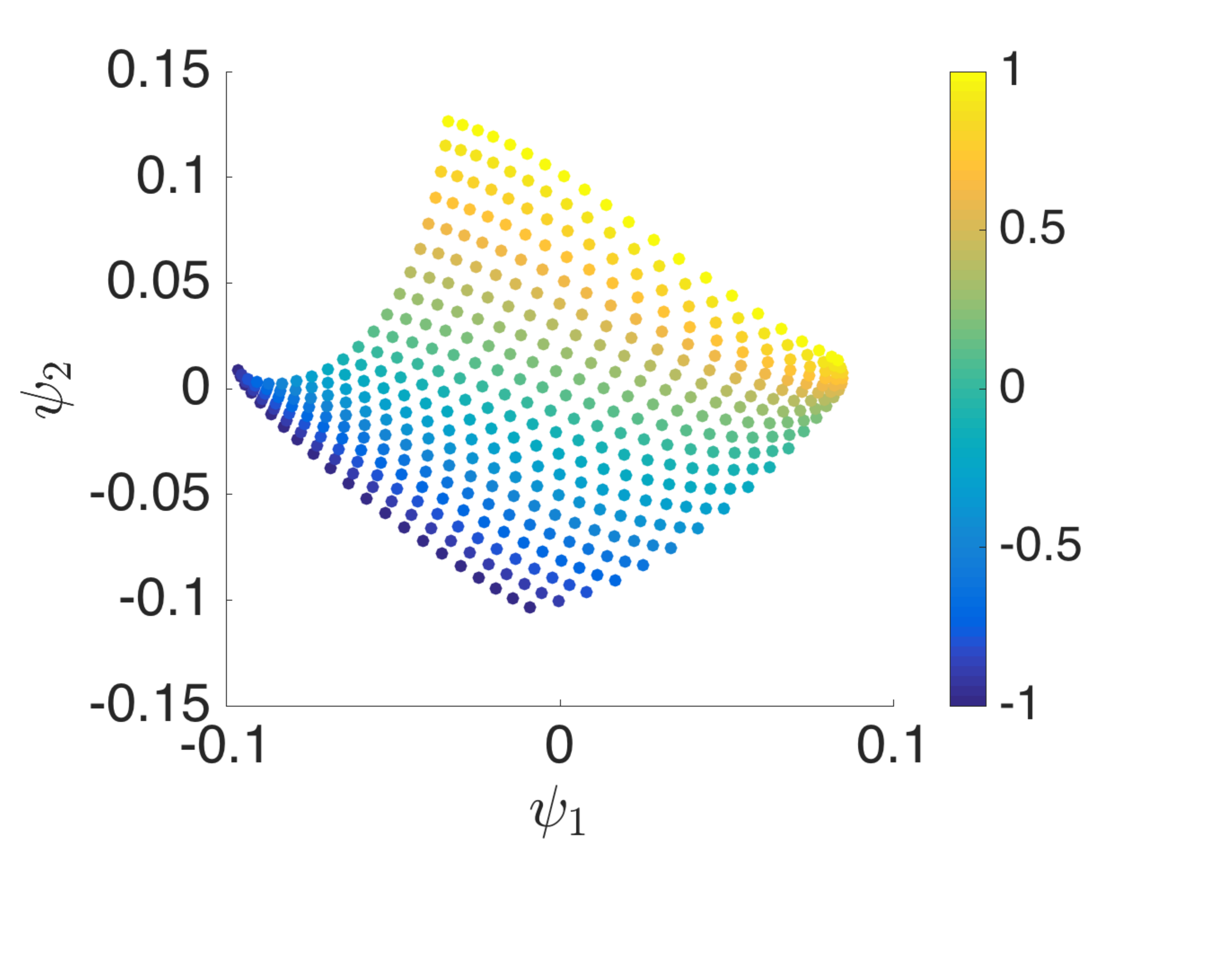}}
\subfloat[]{\includegraphics[height=3.3cm]{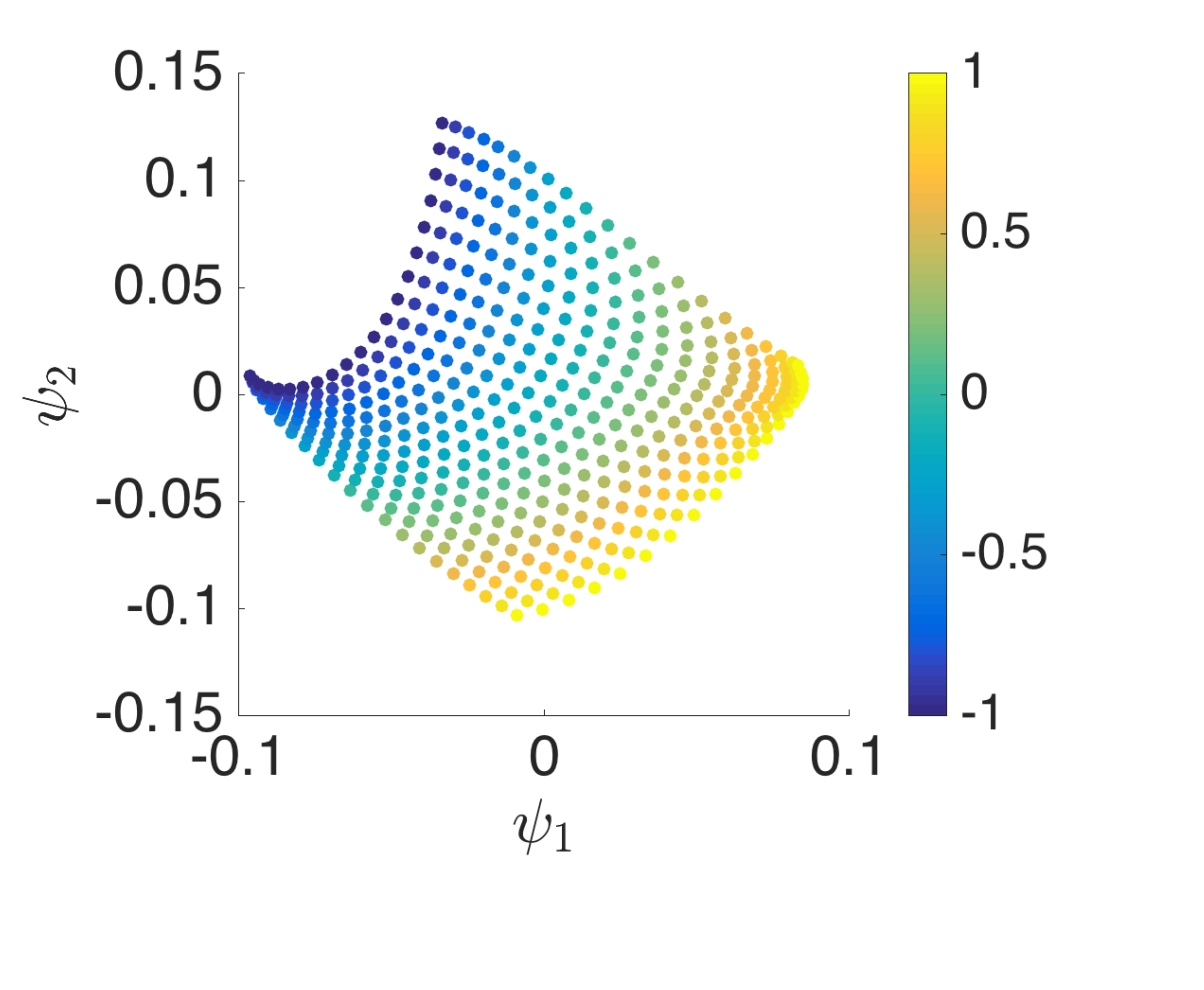}}
\caption{
	(left) Data-driven embedding of the parameters axis of the observations collected from the Bogdanov-Takens system
	(colored according to the true bifurcation map).
	Embeddings built from (a) state variable observations; and (b) observations through a nonlinear invertible function.
    (right) Data-driven embedding of the state variables axis (c) colored by the initial conditions
    of $x_1$, and (d) by the initial conditions of $x_2$.}
\label{fig:BT_results}
\end{figure*}

Our method is applied to the $3$D tensor of trajectories $\mathbf{Y}$ collected from the Bogdanov-Takens system. As described above, $\mathbf{Y}$ consists of (short) trajectories of observations arising from the system initialized with various initial conditions and with various parameters.
We emphasize that the knowledge of the different regimes and the bifurcation
map {\em were not taken into account} in the analysis; only the time-dependent data $\mathbf{Y}$ were considered.

Figure~\ref{fig:BT_results}~(a) depicts the scatter plot of the two
dominant eigenvectors representing the parameters axis. The figure
consists of $N_{p}$ points (the length of the eigenvectors), where
each point corresponds to a single sample $\boldsymbol{y}_{p}\in\mathbb{R}^{N_{v}\times N_{t}}$,
which is associated with a sample $\boldsymbol{p}=(\beta_{1},\beta_{2})$
of parameters values on the $2$D grid depicted in Figure~\ref{fig:BT_params}.
Moreover, each point in Figure~\ref{fig:BT_results}~(a) is colored
by the same color-coding used in Figure~\ref{fig:BT_params}.

We observe that our method discovers an {\em empirical bifurcation
mapping of the system}. Indeed, the obtained representation of the
parameters through the eigenvectors establishes a new coordinate system
with a geometry, built solely from observations, which reflects the
organization of the parameters space according to the true underlying
bifurcation map -- the ``visual homeomorphism'' (stopping short of claiming visual isometry) is clear.

To illustrate the generality of our method, we now apply a {\em nonlinear (yet invertible)} observation function
\[
\boldsymbol{z}(t) = h(\boldsymbol{x}(t))
\]
with $h_k(\boldsymbol{x}(t)) = \sqrt{\boldsymbol{a}_k^T \boldsymbol{x}(t) + \alpha_k}, \ k=1,2$,
where $\boldsymbol{a}_k$ is a random observation vector and $\alpha_k$ is a constant set to guarantee positivity.
Figure~\ref{fig:BT_results}~(b) depicts the scatter plot of the two
dominant eigenvectors representing the parameters axis obtained from the new set of nonlinear observations.
An equivalent organization is clearly achieved.

Figure~\ref{fig:BT_results}~(c) depicts the scatter plot of the two dominant
eigenvectors representing the state variable axis. The plot consists of
$N_{v}$ points (the length of the eigenvectors), where each point
corresponds to a single sample $\boldsymbol{y}_{v}\in\mathbb{R}^{N_{p}\times N_{t}}$,
which is associated with a particular set of initial condition values $\boldsymbol{v}=(y_{1}(0),y_{2}(0))$.
The embedded points are colored in Figure~\ref{fig:BT_results}~(c)
by the initial conditions of the variable $y_1$, and in Figure~\ref{fig:BT_results}~(d)
by the initial conditions of the variable $y_2$.
The color-coding implies that the recovered $2$D space corresponds to the $2$D space of the true variables of the system. In other words, the high dimensional samples $\boldsymbol{y}_{v}$
are embedded in a $2$D space, which recovers a $2$D structure accurately representing the true directions of the hidden, minimal, two
state variables of the system.

\subsection*{Two Coupled Pendula}
\label{sub:CoupledPendulum}

\begin{figure*}[t]
\centering \includegraphics[height=2.7cm]{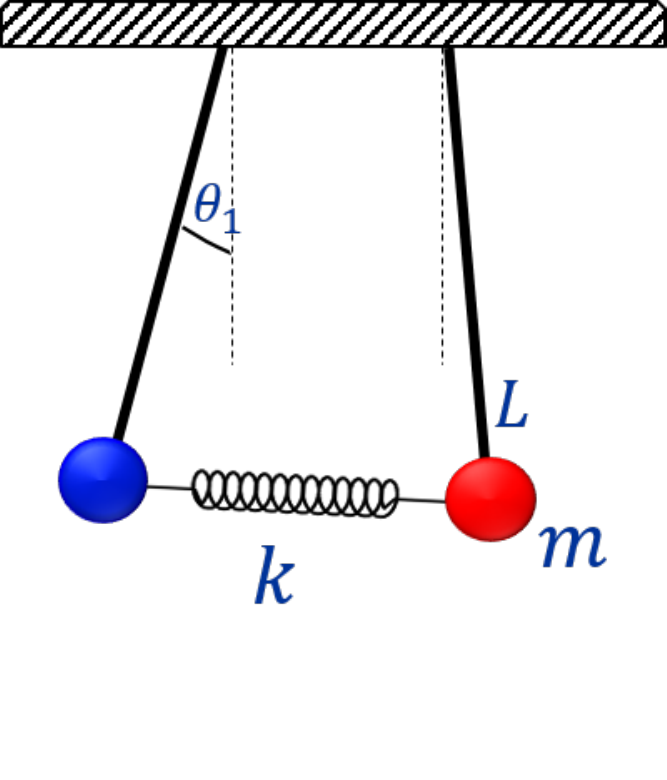}
\centering \includegraphics[height=3cm]{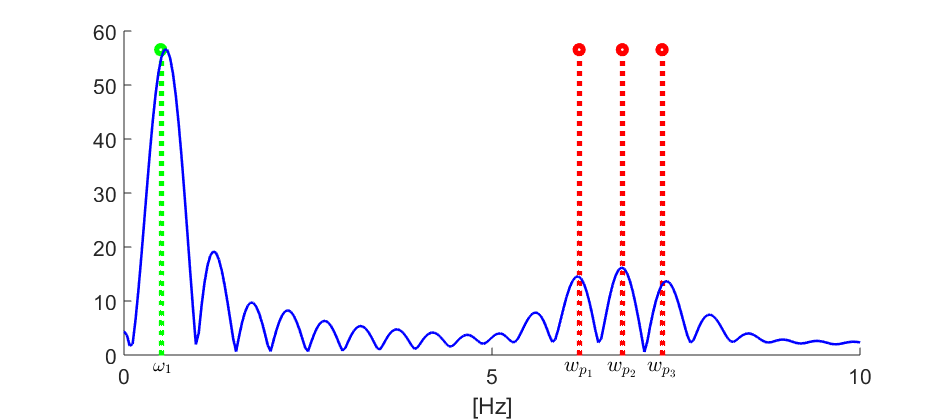}
\centering \includegraphics[height=3cm]{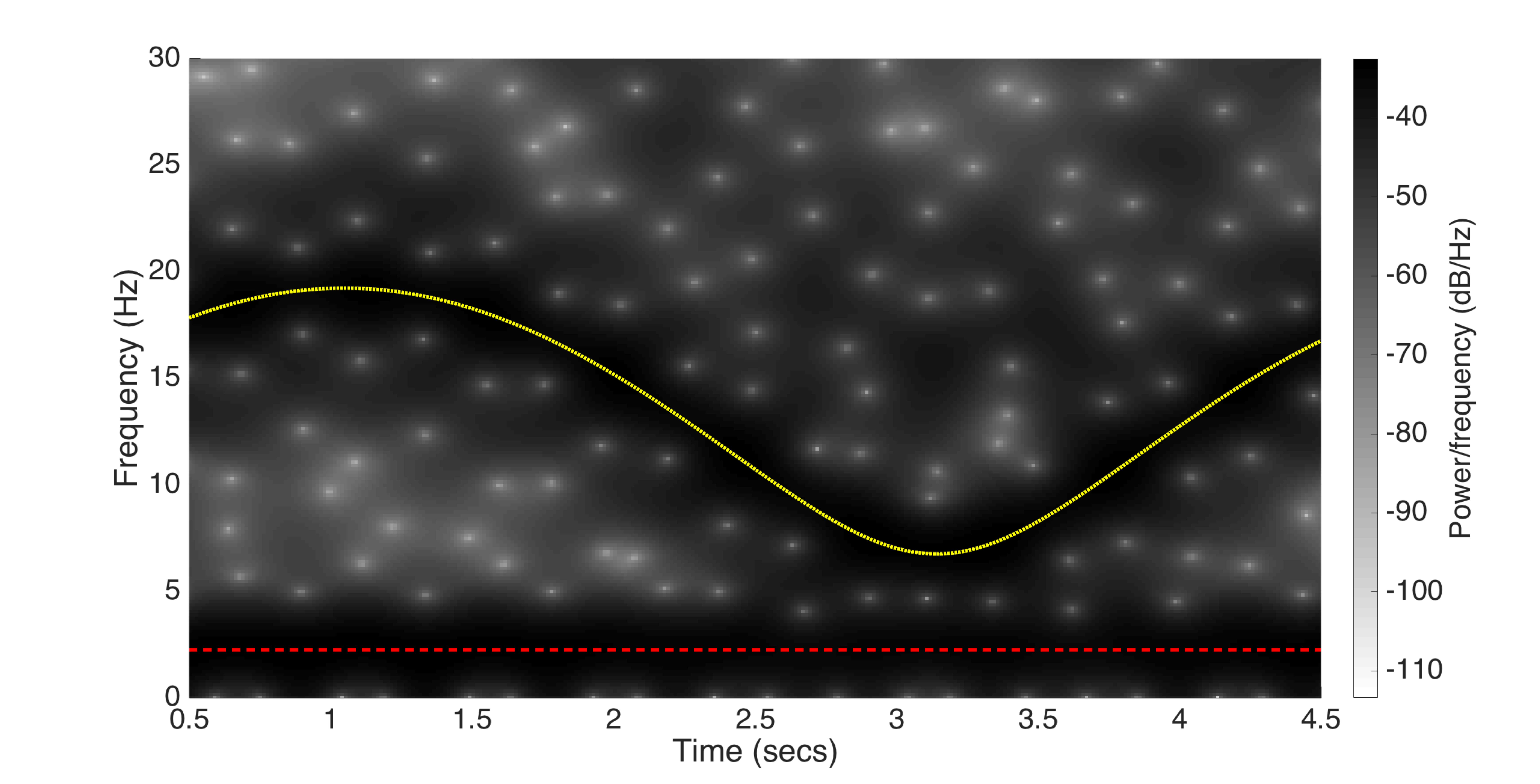}
\caption{(left) Schematic of the two coupled pendula. (middle) The Fourier transform of the principal eigenvector representing
		the time axis (blue) obtained in the first experiment. The dashed lines are the true (hidden) frequencies.
		The dashed green line corresponds to the fixed oscillation
		frequency $\omega_{1}$ and the dashed red lines correspond to
		the varying-across-movies oscillation frequency $\omega_{p}$.
		(right) The Fourier spectrogram of the principal eigenvector representing
the time axis obtained in the second experiment. The two oscillation frequencies $\omega_1$ and $\omega_2(t)$ are overlayed on the spectrogram. The dashed red line corresponds to the fixed oscillation frequency $\omega_{1}$ and the dotted yellow line corresponds to
the time-varying oscillation frequency $\omega_{2}$.}
\label{fig:pendulum}
\end{figure*}

We simulate a system of two simple coupled pendula
with equal lengths $L$ and equal masses $m$, connected
by a spring with constant $k$ as shown in Figure
\ref{fig:pendulum}~(left).
Let $u^{(1)}\left(t\right)$ (resp. $u^{(2)}\left(t\right)$)
denote the horizontal position of the first (resp. second) pendulum, and let $\theta^{(1)}(t)$
and $\theta^{(2)}(t)$ be the angles between each of them and the
vertical axis. Note that a closed-form expression for the positions
cannot be derived for the general case. Yet, in the case of small
perturbations around the equilibrium point, we consider a linearized regime
and assume that $\sin\theta^{(i)} \approx\theta^{(i)}$
for $i=1,2$. 
We present the linear case because it can be trivially reproduced, and thus validated; yet our data-driven approach can equally well
be performed for the more general nonlinear case.
The \ac{ODE} system describing the evolution of
the horizontal positions within this linear regime is given by:
\begin{align}
\begin{cases}
m\ddot{u}^{(1)}=-\frac{mg}{L}u^{(1)}-k\left(u^{(1)}-u^{(2)}\right)\\
m\ddot{u}^{(2)}=-\frac{mg}{L}u^{(2)}+k\left(u^{(1)}-u^{(2)}\right)
\end{cases}\label{eq:ODE}
\end{align}
where $\ddot{u}$ is the second derivative of $u$ and $g$ represents gravity.
To highlight the broad scope of our approach from a data analysis perspective,
we assume that we do not have direct access to the horizontal displacement.
Instead, we generate {\em movies} of the motion of the coupled pendula (observed in the form of
pixels, see below))
in the linear regime. On the one hand, we have a definitive, physical-law based
ground truth described by the solution of the \ac{ODE} of the system (in terms of the two normal modes of the system, see Appendix \ref{sec:SI_Coupled_Pendulum} for details).
On the other hand, we only have access to {\em high-dimensional nonlinear
observations} of the system: the pixels of the frames of the movie.

In the first experiment, we generate {\em three} movies, corresponding to three different spring constants $k_p$, where $p=1,2,3$.
As a consequence, the motion in each movie consists of two oscillations:
\begin{equation*}
	\omega_{1}=\sqrt{\frac{g}{L}}, \quad \omega_{p}=\sqrt{\frac{g}{L}+\frac{2k_{p}}{m}}
\end{equation*}
where $L = 1$, $m = 1$, $g = 9.8$ and $k_{p}\in \left\{750, 900, 1050 \right\}$.
Each movie comprises $400$ frames and is $4$ seconds long. 
Each frame of the movie consists of $20 \times 40$ black, white, red, green and blue pixels;
An example of one movie can be found here: https://youtu.be/rjUxeT-ShQc
Our data is therefore $\boldsymbol{Y}\in\mathbb{R}^{N_{p}\times N_{v}\times N_{t}}$
where $N_p = 3$, $N_v = 800$ and $N_t = 400$.
We apply Algorithm \ref{algo1} to the input data $\mathbf{Y}$ (we could apply it to the digitization
of an actual experiment).
Figure \ref{fig:pendulum} (middle) displays the Fourier transform of
the most dominant eigenvector from the time embedding.
It is clearly seen that all the frequencies $\omega_1$ and $\omega_p$ are captured, 
where the fixed frequency $\omega_1$ is dominant and the varying-across-movies frequencies $\omega_p$ are less pronounced, as expected.
This experiment demonstrates the features of our methodology, and the empirical results are presented in light of the known ground truth. 

In the second experiment, we simulate a system with {\em time-varying spring constant} $k(t)$.
In this case, we do not have an analytic solution for any $k(t)$
(See Appendix \ref{sec:SI_Coupled_Pendulum}), and therefore, there is no definitive ground truth; yet, we demonstrate that our method enables us to recover two normal modes even in this (explicitly unsolvable) scenario.
Now, we generate a movie, where the spring constant
$k(t)$ changes over time according to $k(t)=1000+800\sin(1.5t)$.
This movie is available in the following link: https://www.youtube.com/watch?v=I8C6Yt3b2tk.

In the context of our data processing, the variables axis consists of the 800 pixels, and the time axis represents the 400 frames.
To simplify the exposition, we focus here on a {\em single} parameter setting, so that the parameters axis is degenerate,
i.e., $N _t = 400$, $N_v = 800$, $N_p = 1$, and $\mathbf{Y}\in\mathbb{R}^{800\times400}$.

We apply our method to the data from the simulated movie. Figure \ref{fig:pendulum} (right)
depicts the Fourier spectrogram of the principal eigenvector representing
the time axis, obtained by diffusion maps {\em with the informed metric} after $1$ iteration.
We observe that the empirical representation of the time axis identifies
two oscillation frequencies $\omega_{1}=\sqrt{g/L}$ and $\omega_{2}(t) = \sqrt{g/L+2k(t)/m}$, which are emphasized by overlayed curves.
 
\begin{figure}[t]
\centering
\includegraphics[width=0.5\columnwidth]{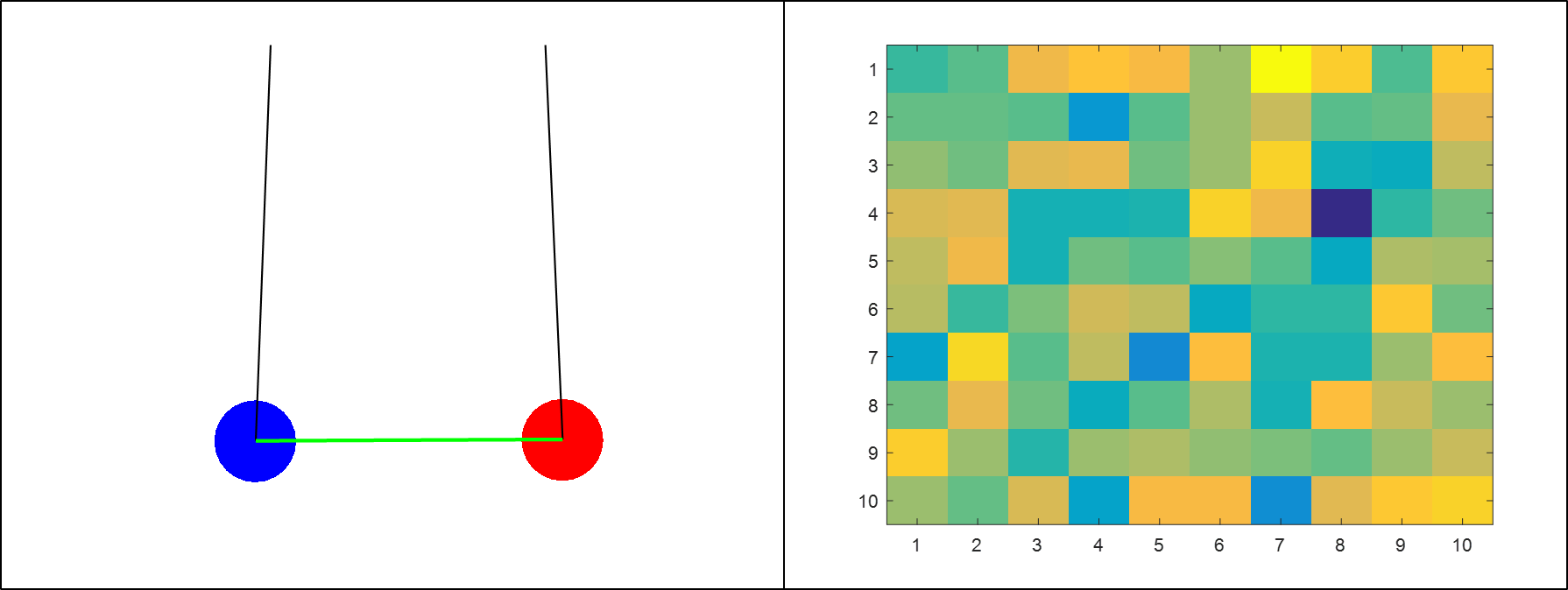}
\includegraphics[width=0.5\columnwidth]{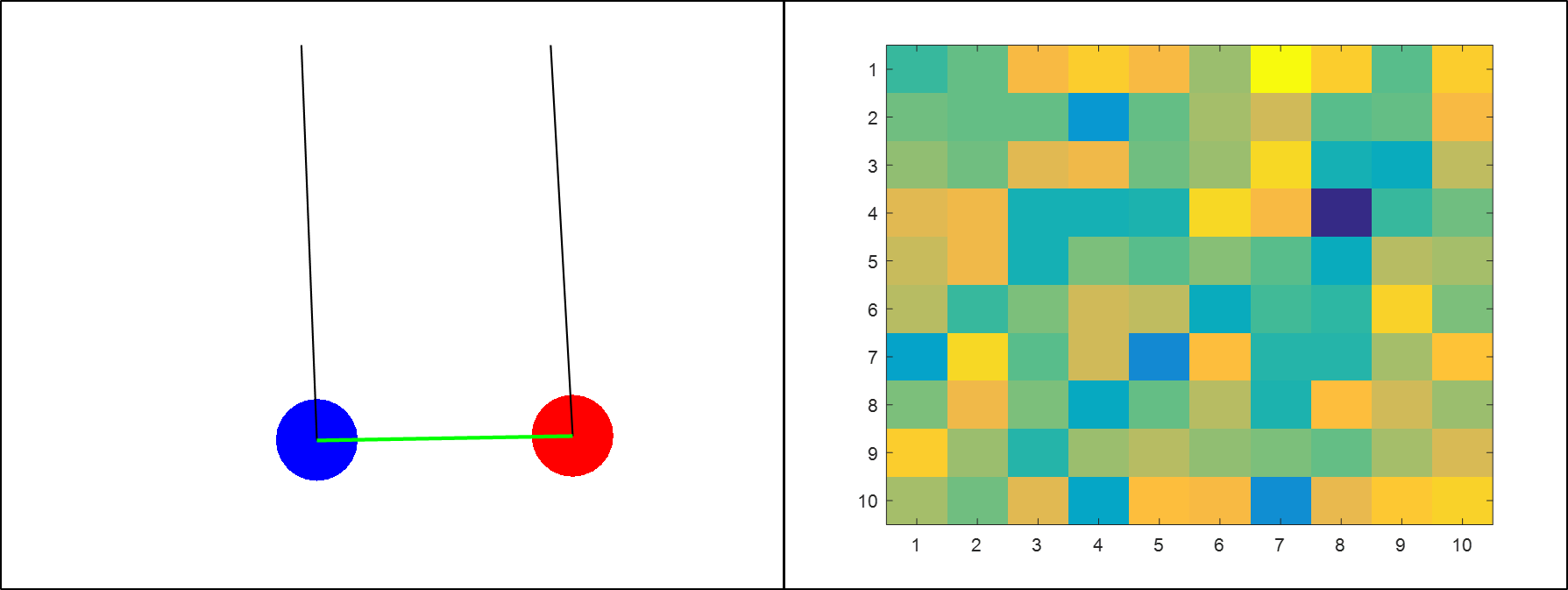}
\includegraphics[width=0.5\columnwidth]{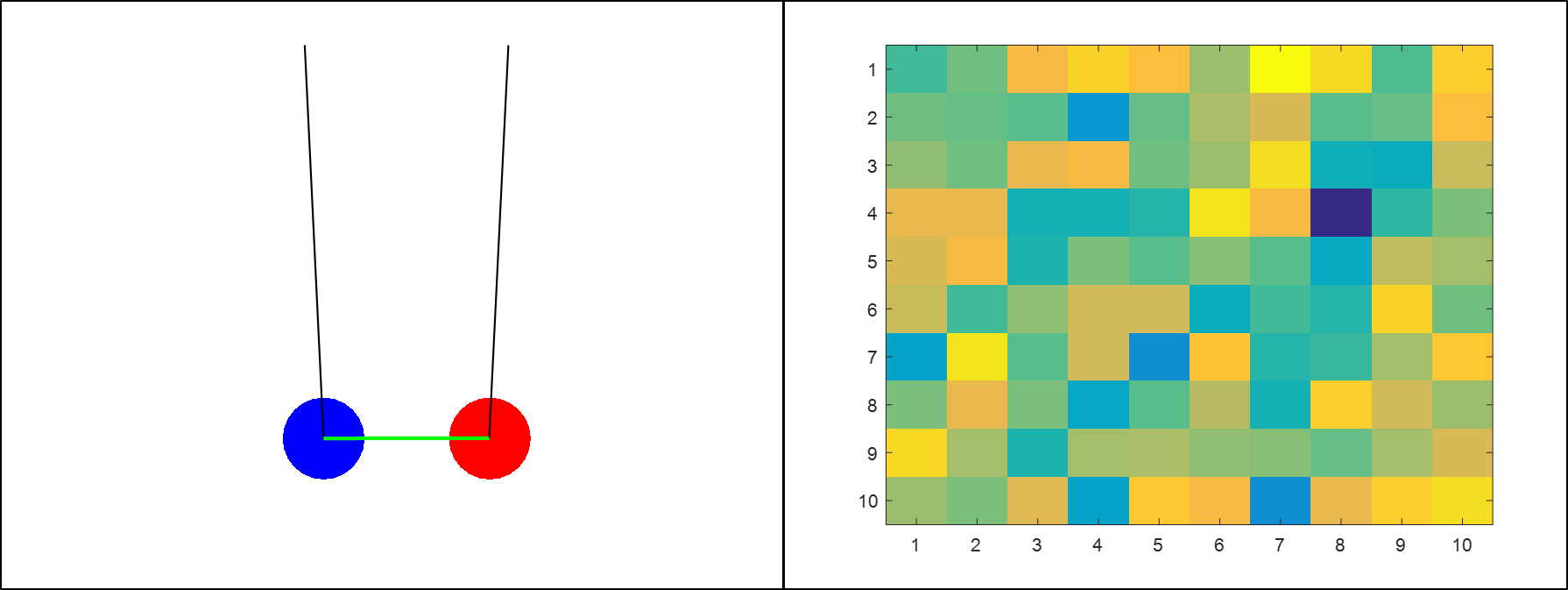}
\caption{An example of $3$ snapshots of the coupled pendulum system paired with their random projection counterparts.}
\label{fig:snapshots}
\end{figure}

To highlight the scope and potential of our approach we now apply a {\em fixed, invertible, random projection} to each frame of the movie. In other words, each frame of the movie was multiplied by a fixed matrix, whose columns were independently sampled from a multivariate normal distribution and normalized to have a unit norm. The resulting movie with the projected frames can be found in the following link: https://www.youtube.com/watch?v=xz0hzQTyPGo.
Figure \ref{fig:snapshots} depicts an example of $3$ snapshots of the coupled pendulum system paired with their random projection counterparts.
The obtained parametrization, representing the time axis obtained from the new, ``scrambled'' movie, is presented in Figure \ref{fig:pendulum_results_rp}, where we observe that the same two frequencies $\omega_1$ and $\omega_2(t)$ are captured by our method, despite the additional, {\em unknown} ``observation'' function.
An additional experiment appears in Appendix \ref{sec:SI_Coupled_Pendulum}.

\begin{figure}[t]
\centering \includegraphics[height=3.5cm]{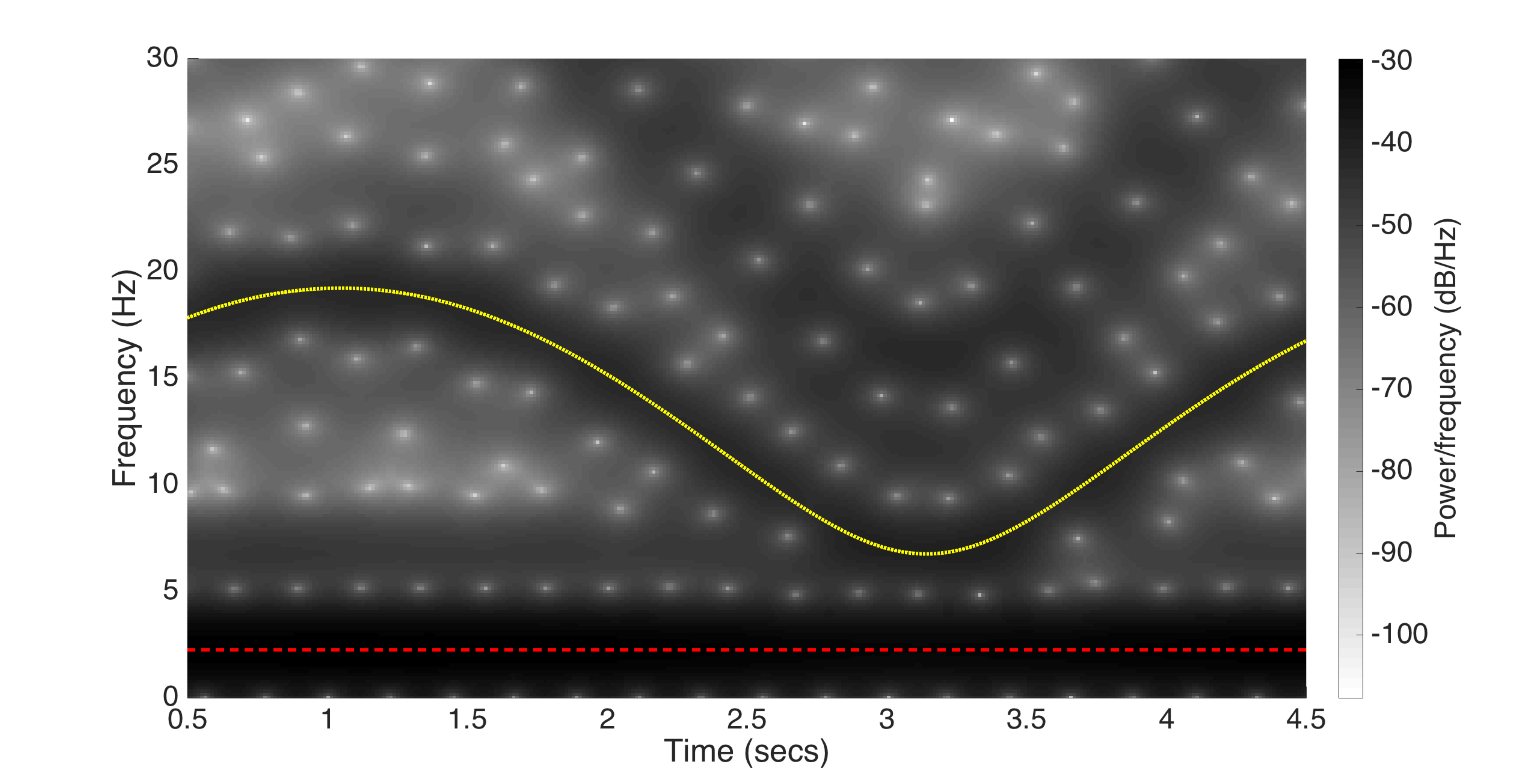}
\protect\caption{The Fourier spectrogram of the principal eigenvector representing
the time axis. These results are based on the random projections of the movies frames with the same time-varying spring constant.
The two frequencies $\omega_{1}$ and $\omega_{2}(t)$ are overlayed on the spectrogram. 
The dashed red line corresponds to the fixed oscillation
frequency $\omega_{1}$ and the dotted yellow line corresponds to
the time-varying oscillation frequency $\omega_{2}$.}
\label{fig:pendulum_results_rp}
\end{figure}

\section*{Summary and Discussion}

Obtaining predictive dynamical equations from data lies at the heart of science and engineering modeling, and is the linchpin of our technology. 
Today we witness the development of mathematical techniques that operate directly on observations, i.e., data-driven, and appear to circumvent the need to ``manually'' select variables and parameters and to derive accurate equations.
The core of this new methodology has shifted to the development of the mathematics and algorithms that systematically transition from data to the analysis of the model, and thus to making predictions, without ever deriving the model in closed-form.
Our work here presents an illustration of this path, in an attempt to extend classical transformative results of linear system identification \cite{kalman1963mathematical} and realization theory \cite{moore1981principal} (see also \cite{lall1999empirical}). 
It is important to note that we assumed we had {\em plenty} of data, whereas the question about the fewest measurements required for modeling, and how to efficiently plan and conduct their sampling, has been left open for future work.

We have demonstrated that (invertible) functions of our measurements give rise to homeomorphic (and possibly isometric) embeddings, 
thereby conveying the same prototypical behaviour (in the spirit of Whitney and the even stronger Nash embedding theorems,
and more specifically Takens embedding for dynamical systems
\cite{packard1980geometry,whitney1944self,takens1981detecting,nash1956imbedding}).
This points towards the feasibility of ``gauge invariant'' data mining \cite{Singer2008,talmon2013empirical,dsilva2013nonlinear,Talmon2014,dsilva2016data}, through algorithms that do not depend on the measurement modality. 

While we only use indices for the trials and the measurement channels, we did keep here the sequention parametrization of the measurements by time;
this is {\em not necessary}, and one could use only labels also in the time axis (denoting what measurements were obtained at the same moment
in time, but without knowing the actual time). This would lead to the correct ordering of the measurement snapshots,
without providing actual time stamps for them \cite{dsilva2015temporal}. 
More generally, semi-supervised learning and ``manifold completion'' tools can be used to fruitfully
fill in missing data, interpolate and (modestly) extrapolate the input-output functions learned here.
And while our approach has been built to directly apply to real world data, it would be particularly
exciting to apply it as a second, ``debriefing'' layer to create minimal, and hopefully balanced, realization of
the very high dimensional representations of such data learned by a first reconnaissance by deep nets \cite{dsilva2015temporal}.

In learning features of an input-output description, we provided the algorithm with data we knew were ``relevant''.  How would the algorithm perform if we presented it with some extraneous data,
along with the relevant ones? This subject, which is crucial for learning is addressed in \cite{lederman2015learning}, 
where data that can be deduced to be measurements of the same process can be systematically identified.
It is also clear that the conditions for convergence of our ``triple successive iteration'' scheme
(what one might call a ``consistent input output balancing/renormalization'') should be 
established mathematically.
To this end, the seminal work of Breiman \cite{breiman1985estimating} (and also \cite{chi2016convex}) will clearly be useful in this effort, while also laying the ground for statistical and possibly Bayesian interpretations.

While our construction takes into account the coupling between state space, parameters and time, our final representations for
each ``entity'' are separate. Perhaps the most natural extension of this work lies in the direction of detecting changes in the state or parameter {\em embedding dimensions}.
Whether distinct (regular perturbations) or joint (singular perturbations), these dimension
changes are crucial for successful model reduction. For this purpose, in future work, we will examine the expansion of the high-dimensional data in
the joint parameters-state-time space obtained by a product of the inferred representation components.
Thus establishing causality and obtaining an intrinsic form of the governing equations of the dynamical system.

\section{Appendix: Diffusion Maps} \label{Sec:Diffuion_Maps}

One class of data-driven methods for analyzing complex data sets is manifold learning.
The main assumption in manifold learning is that often data are
constrained to lie on or around a low dimensional manifold. The idea
is then to build a low dimensional embedding of the data in a
data-driven way, such that it parametrizes the underlying manifold structure.
A particular manifold learning method, diffusion maps, is used as a building block in our approach and is briefly described next.

Given a set $\{\boldsymbol{y}_{t}\}$ of $N_{t}$ samples of observations of the system, let $\boldsymbol{W}$ be an $N_{t}\times N_{t}$
pairwise affinity matrix, whose $(t,s)$-th
entry is defined by
\begin{equation}
W_{t,s}=e^{-\|\boldsymbol{y}_{t}-\boldsymbol{y}_{s}\|^{2}/\epsilon}
\label{eq:affinity_matrix}
\end{equation}
where $\epsilon>0$ is a positive scale parameter. Common practice
is to interpret the set of samples and the affinity matrix as a graph,
where the samples are the graph nodes and the affinity matrix determines
the weights of the edges, i.e., node $\boldsymbol{y}_{t}$ is connected
to node $\boldsymbol{y}_{s}$ by an edge with weight $W_{t,s}$.
Setting the appropriate value of the kernel scale $\epsilon$ has been the subject of many studies, e.g., \cite{hein2005intrinsic,coifman2008graph}. One standard approach is to set it as the median of all the distances $\|\boldsymbol{y}_{t}-\boldsymbol{y}_{s}\|^{2}$ in $W_{t,s}$; in the graph interpretation, this results in well connected graphs, and therefore, enables the user to take into account a large number of relationships between the available samples.

The next step aggregates the pairwise affinities/graph connections
into a global parametrization. Prior to that, the affinity matrix
is traditionally normalized; several normalizations are considered
in the literature, and here we present a particularly common one. Let $\boldsymbol{D}$ be a diagonal
matrix whose main diagonal comprises the sum of rows of $\boldsymbol{W}$, and let $\boldsymbol{A}=\boldsymbol{D}^{-1}\boldsymbol{W}$
be a row-stochastic matrix.
In the graph interpretation, $\boldsymbol{A}$ can be seen as a transition
probability matrix defining a Markov chain on the graph, where $A_{t,s}$
is the probability to ``jump'' from node $\boldsymbol{y}_{t}$ to node
$\boldsymbol{y}_{s}$ in one Markov chain step.

The global parametrization is obtained by applying the eigenvalue
decomposition (EVD) to $\boldsymbol{A}$. Let $\lambda_{\ell}$ denote
the eigenvalues, ordered in decreasing order, and let $\boldsymbol{\psi}_{\ell}$
denote the corresponding (right) eigenvectors. Note that $\boldsymbol{A}$ is row stochastic, and hence its largest eigenvalue is $\lambda_0=1$, corresponding to an all-ones trivial eigenvector $\boldsymbol{\psi}_0$; since both do not carry information on the data, they are typically ignored.
The diffusion maps embedding of the samples is defined as the following
nonlinear map for some $\tau>0$:
\begin{equation}\label{eq:dmaps}
\boldsymbol{y}_{t}\mapsto(\lambda_{1}^{\tau}\psi_{1}(t),\lambda_{2}^{\tau}\psi_{2}(t),\ldots,\lambda_{m}^{\tau}\psi_{m}(t))
\end{equation}
where each sample $\boldsymbol{y}_{t}$ is embedded in $\mathbb{R}^m$ by the values of the $m$ eigenvectors associated with the $m$ largest eigenvalues.

The diffusion maps embedding bears two important properties.
First, it can be shown that the Euclidean
distance between the {\em embedded} points approximates the diffusion distance,
a distance defined by the induced transition probabilities and is
closely related to the geodesic distance on the assumed underlying
manifold \cite{Coifman2006}. Second, the eigenvectors $\boldsymbol{\psi}_{\ell}$ form
an orthonormal basis for any real function defined on the sample set
$\{\boldsymbol{y}_{t}\}$.
For more details, see \cite{coifman2005geometric,Coifman2006}.

\section{Appendix: Informed Metric Construction} \label{appx:informed_metric}

\subsection*{Partition Trees}

We build a partition tree for each axis based on a given metric.
Each partition tree is composed of $L+1$ levels, where a partition $\mathcal{I}_{l}$
of the samples is defined for each level $0\leq l\leq L$. The partition
$\mathcal{I}_{l}$ at level $l$ consists of $n(l)$ mutually disjoint
non-empty subsets of samples, termed folders and denoted by $I_{l,i}$,
$i\in\{1,...,n(l)\}$, i.e.,:
\begin{equation}
\mathcal{I}_{l}=\{I_{l,1},I_{l,2},...,I_{l,n(l)}\}.
\end{equation}
The partition tree has the following properties:
\begin{itemize}
\item The finest partition ($l=0$) is composed of singleton folders, termed
the ``leaves'', where $I_{0,i}=\{\boldsymbol{v}_i\}$ and the number of leaves
is the total number of samples.
\item The coarsest partition ($l=L$) is composed of a single folder $I_{L,1}$,
which is termed the ``root'' of the tree and consists of all the samples.
\item Each folder at level $l-1$ is a subset of a folder from level $l$,
i.e., the partitions are nested such that if $I\in\mathcal{I}_{l}$,
then $I\subseteq J$ for some $J\in\mathcal{I}_{l+1}$.
\end{itemize}
The partition tree is the set of all folders at all levels
\begin{equation}
\mathcal{T}=\{I_{l,i}\;\vert\;0\leq l\leq L,\;1\leq i\leq n(l)\}.
\end{equation}
See \cite{mishne2015hierarchical} for more details.

There are multiple ways to build such partition trees.
The different construction methods can be divided into two classes:
bottom-up construction and top-down construction. Broadly, a bottom-up
construction begins with the definition of the lower levels, initially
by grouping the leaves/samples, e.g., using k-means.
Then, these groups are further grouped in an iterative procedure to
create the next levels, ending at the root, in which all the samples are placed under a single folder.
A top-down construction is typically
implemented by an iterative clustering method, initially applied to
the entire set of samples, then refined over the course of the iterations,
starting with the root of the tree and ending at the leaves.

\subsection*{Iterative Metric Construction}

The construction of the partition tree described above relies on a metric between the samples.
We propose a procedure, in which the construction of the tree relies on an iteratively evolving ``informed metric'' induced by partition trees on the coordinates of the samples.
Namely, the construction of $\mathcal{T}_{v}$ relies on a
metric between the samples $\boldsymbol{y}_{v}$ and the construction
of $\mathcal{T}_{t}$ relies on a metric between the samples $\boldsymbol{y}_{t}$.
Given $\mathcal{T}_{v}$ and $\mathcal{T}_{t}$, the informed metric between the samples $\boldsymbol{y}_{p}$
is constructed, and then, used to build a partition tree $\mathcal{T}_{p}$ of the samples
$\boldsymbol{y}_{p}$. In the second substep within the iteration, $\mathcal{T}_{p}$ can be used to construct refined metrics
between $\boldsymbol{y}_{v}$ and between $\boldsymbol{y}_{t}$. In what follows we describe one full iteration in detail.

Let $\mathcal{T}_{v}$ and $\mathcal{T}_{t}$ denote finite partition trees of the samples $\left\{ \boldsymbol{y}_{v}\right\} $ and $\left\{ \boldsymbol{y}_{t}\right\} $, respectively.
The partition trees $\mathcal{T}_{v}$ and $\mathcal{T}_{t}$ induce
a multiscale decomposition on the data, particularly, {\em of the coordinates
of the samples} $\{ \boldsymbol{y}_{p} \}$.
Here, we show how this decomposition is used to construct an informed metric between the samples $\{ \boldsymbol{y}_{p} \}$, and in turn, a partition tree $\mathcal{T}_p$ on $\{\boldsymbol{y}_{p}\}$; we formulate it by the construction
of a {\em data-adaptive filter bank}. Define the filter $g_{I\times J}$
for each $I\in\mathcal{T}_{v}$ and $J\in\mathcal{T}_{t}$ by:
\begin{equation}
g_{I\times J}=\frac{w(I,J)}{\vert I\vert\vert J\vert}\mathds{1}_{I}\otimes\mathds{1}_{J},
\label{eq:filter2d}
\end{equation}
where $w(I,J)$ are positive weights, $| \cdot |$ denotes the cardinality of a set, $\otimes$ denotes the outer product, and $\mathds{1}_{I}$ and $\mathds{1}_{J}$ denote the indicator vectors of the samples in folder $I\in\mathcal{T}_{v}$ and in folder $J\in\mathcal{T}_{t}$.
For each filter we calculate the inner product between the filter $g_{I\times J}$ induced by folders
$I$ and $J$ and a sample $\boldsymbol{y}_{p}$, yielding a scalar
coefficient $f_{I\times J}$:
\begin{equation}
\begin{split}f_{I\times J}(\boldsymbol{y}_{p}) & =\langle g_{I\times J},\boldsymbol{y}_{p}\rangle\\
 & =\frac{w(I,J)}{\vert I\vert\vert J\vert}\sum_{\boldsymbol{v}\in I,t\in J}Y(\boldsymbol{p},\boldsymbol{v},t)
\end{split}
\label{eq:coef}
\end{equation}
By definition, $f_{I\times J}(\boldsymbol{y}_{p})$ can be viewed as a nonlocal, informed low-pass filter, since it computes a nonlocal average of the coordinates of $\boldsymbol{y}_{p}$ prescribed by the folders $I$ and $J$.
The transform $\mathcal{F}_{\mathcal{P}}(\boldsymbol{y}_{p})$ is then defined by
\[
\mathcal{F}_{\mathcal{P}}(\boldsymbol{y}_{p})=\left\{ f_{I\times J}(\boldsymbol{y}_{p})|\forall I\in\mathcal{T}_{v},\forall J\in\mathcal{T}_{t}\right\}.
\]
In other words, $\mathcal{F}_{\mathcal{P}}(\boldsymbol{y}_{p})$ is the collection of
all coefficients $f_{I\times J}(\boldsymbol{y}_{p})$ resulting from
the applications of all possible filters $g_{I\times J}$ to $\boldsymbol{y}_{p}$.

The main property of $\mathcal{F}_{\mathcal{P}}$ stems from the utilization of the co-dependencies of the variables
and time samples in a data-driven manner; by combining mutually related, yet nonlocal, sets of variables and time samples, an {\em informed} representation of the samples from the parameters axis is obtained.
In addition, $\mathcal{F}_{\mathcal{P}}(\boldsymbol{y}_p)$ can be viewed as appending coordinates to $\boldsymbol{y}_p$ by computing averages according to the partition trees.
A particular property of the $\ell_1$ norm of these ``vectors'' $\mathcal{F}_{\mathcal{P}}(\boldsymbol{y}_p)$ is described next.

We set the weights as
\begin{equation}
w(I,J)=\left(\frac{\vert I\vert}{N_{v}}\right)^{\beta_{v}+1}\left(\frac{\vert J\vert}{N_{t}}\right)^{\beta_{t}+1},\label{eq:omega2d}
\end{equation}
where $\beta_{v}$ weighs the bi-folder $I\times J$ based on the
relative size of folder $I\in\mathcal{T}_{v}$ and $\beta_{t}$ weighs
the bi-folder based on the relative size of $J\in\mathcal{T}_{t}$.
These values should be set according to the smoothness of the data
along the respective axis; the smoother the data are along each axis,
the larger the corresponding coefficient should be, thereby promoting larger folders (coarser representation).
With these weights, the $\ell_{1}$ distance between the transformations,
i.e.,:
\begin{equation}
\Vert \mathcal{F}_{\mathcal{P}}(\boldsymbol{y}_{p})-\mathcal{F}_{\mathcal{P}}(\boldsymbol{y}_{q})\Vert_{1}.
\end{equation}
equals the \ac{EMD} between the samples
$\boldsymbol{y}_{p}$ \cite{mishne2015hierarchical}.
By definition, we have
\begin{equation}
\Vert \mathcal{F}_{\mathcal{P}}(\boldsymbol{y}_{p})-\mathcal{F}_{\mathcal{P}}(\boldsymbol{y}_{q})\Vert_{1} = \sum _{I \in \mathcal{T}_v, J \in \mathcal{T}_t} | f_{I\times J}(\boldsymbol{y}_{p}) - f_{I\times J}(\boldsymbol{y}_{q})|
\label{eq:transform}
\end{equation}
This distance thus enables us, in an efficient
data-driven manner, to compute the \ac{EMD} based only on nonlocal
averaging filters induced by the partition trees.
Once the metric is constructed, it can be used to build a partition tree $\mathcal{T}_p$ on the samples $\boldsymbol{y}_p$.

The construction of the informed metric between the samples $\boldsymbol{y}_{p}$
described above is repeated in an analogous manner to build informed
metrics between the samples $\boldsymbol{y}_{v}$ and between the
samples $\boldsymbol{y}_{t}$. The entire iterative construction procedure
is summarized in Algorithm \ref{algo2}.
Proving convergence for this ``iterative, self-consistent renormalization'' of the coordinates,
is the subject of current research.

Finally, we remark on two possible variants of the algorithm.
First, in the current presentation, the construction of the tree is carried out by
relying directly on the {\em informed} \ac{EMD}.
Alternatively, we can first apply diffusion maps to the samples based on the \ac{EMD},
and then use the diffusion distances to build the trees.
This will introduce an additional, nonlinear stage,
in which the local information conveyed by the \ac{EMD} is aggregated into a more global
metric between the samples (realized by appending additional ``informed'' coordinates to
the samples).
The algorithm outline for this variant is given in Algorithm \ref{algo1}.
A second variant relates to the definition of the filters $g_{I \times J}$, which is based on indicator functions $\mathds{1}_{I}$ (tensor Haar basis) on the partition tree folders $I$. Instead, we can use the eigenvectors stemming from diffusion maps.
Specifically, let $\boldsymbol{\psi}_{\ell}^{\mathcal{V}}$ and $\boldsymbol{\psi}_{\ell'}^{\mathcal{T}}$ be two eigenvectors resulting from diffusion maps applied (separately) to $\{\boldsymbol{y}_v\}$ and $\{ \boldsymbol{y} _t \}$, respectively.
Then, an alternative definition of the filters in \eqref{eq:filter2d} can be
\begin{equation}
g_{\ell,\ell'}=\boldsymbol{\psi}_{\ell}^{\mathcal{V}} \otimes \boldsymbol{\psi}_{\ell'}^{\mathcal{T}}.
\end{equation}
These two variants will be further explored in future work.

\section{Appendix: More on the Two Coupled Pendula (Autonomous and non-Autonomous)}
\label{sec:SI_Coupled_Pendulum}

\subsection*{Time-varying spring constant}

Our two coupled pendula illustration was chosen so as to provide an easily verifiable ``ground truth'' example in the
linearized regime, and do so involving not only {\em static} but also {\em time-dependent}
inputs.
However, even in this linearized regime, when the spring constant varies in time, the system is not solvable in closed form.
The evolution equations read:
\begin{align}
\begin{cases}
m\ddot{u}^{(1)}=-\frac{mg}{L}u^{(1)}-k(t)\left(u^{(2)}-u^{(1)}\right)\\
m\ddot{u}^{(2)}=-\frac{mg}{L}u^{(2)}+k(t)\left(u^{(2)}-u^{(1)}\right).
\end{cases}\label{eq:ODE2}
\end{align}
In the typical constant $k$ scenario the system is described by two ``normal modes'',  which are obtained by rotating the coordinate system by $45^\circ$ leading to two independent \acp{ODE}:
\begin{align}
\begin{cases}
\ddot{v}^{(1)}=-\omega_1^2v^{(1)} \\
\ddot{v}^{(2)}=-\omega_2^2(t) v^{(2)}
\end{cases}\label{eq:ODE3}
\end{align}
where $v^{(1)} \triangleq u^{(1)} + u^{(2)}$ and $v^{(2)} \triangleq u^{(1)} - u^{(2)}$, and
\begin{equation*}
	\omega_{1}=\sqrt{\frac{g}{L}}, \quad \omega_{2}(t)=\sqrt{\frac{g}{L}+\frac{2k(t)}{m}}
\end{equation*}
The solution of the first differential equation establishes the first normal mode: a solution with a single oscillation frequency $v^{(1)}(t) = A_1 \cos(\omega_1 t)$,
where $A_1$ is a constant depending on the initial conditions.
The second equation is of the form of a Schr\"{o}dinger equation with time-varying potential, $\ddot{v}^{(2)}+\omega_2^2(t) v^{(2)}=0$, and there is no known analytical solution method in the current literature. Yet, several numerical methods exist, e.g., \cite{bremer2015improved}.

If we assume a constant spring, then the second equation becomes linear and similar to the first.
Specifically, for the following initial conditions:
\[
\dot{u}^{(1)}\left(0\right)=0,u^{(1)}\left(0\right)=\delta,\dot{u}^{(2)}\left(0\right)=0,u^{(2)}\left(0\right)=0
\]
where $\dot{u}$ is the first derivative of $u$, and $0<\delta\in\mathbb{R}$
is assumed to be sufficiently small to make the linearization an acceptable approximation,
the closed-form solution of the \ac{ODE} in \eqref{eq:ODE} consists of
a linear combination of two normal modes (oscillations) and is given
by:
\begin{equation}
\begin{cases}
u^{(1)}\left(t\right)=\frac{1}{2}\delta\cos\left(\omega_{1}t\right)+\frac{1}{2}\delta\cos\left(\omega_{2}t\right)\\
u^{(2)}\left(t\right)=\frac{1}{2}\delta\cos\left(\omega_{1}t\right)-\frac{1}{2}\delta\cos\left(\omega_{2}t\right)
\end{cases}\label{eq:ODE_Solution}
\end{equation}
where
\[
\omega_{1}=\sqrt{\frac{g}{L}},\ \ \omega_{2}=\sqrt{\frac{g}{L}+\frac{2k}{m}}.
\]

\subsection*{An Additional Experiment}

\begin{figure}[t]
\centering \includegraphics[height=3cm]{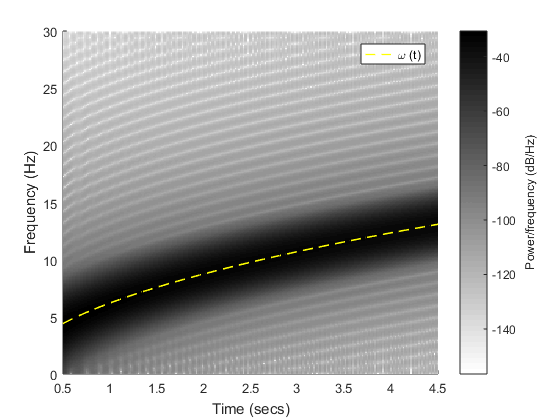}
\centering \includegraphics[height=3cm]{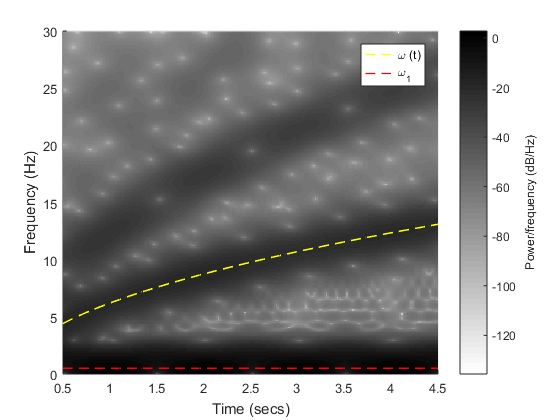}
\caption{(top) The Fourier spectrogram of the numerical solution of $v^{(2)}$ in \eqref{eq:ODE3}. (bottom) The Fourier spectrogram of the principal eigenvector representing the time axis. 
The two frequencies $\omega_{1}$ and $\omega_{2}(t)$ are overlayed on the spectrograms. 
The dashed red line corresponds to the fixed oscillation
frequency $\omega_{1}$ and the dotted yellow line corresponds to
the time-varying oscillation frequency $\omega_{2}$.}
\label{fig:pendulum_results_rp}
\end{figure}

We repeat the experiment reported in the main text.
Here, to help bridging the gap between the two experiment variants: one with a fixed spring constant and verifiable ground truth, and the other with time-varying spring constant without known definitive analytic foundation, we simulate a system with a {\em particular time-varying spring constant}, so that the system has known numerical solution.
The simulated time-varying spring constant used here is: $k(t) = 1500*t + 10$. The remaining details of the simulation is left unchanged and is described in the main text.

The empirical solution of $v^{(2)}$ in \eqref{eq:ODE3} is displayed in Figure \ref{fig:pendulum_results_rp}(top). The results of the application of Algorithm \ref{algo1} are presented in Figure \ref{fig:pendulum_results_rp}(bottom), where the Fourier spectrogram of the principal eigenvector. 
Here as well, we observe that the same two frequencies $\omega_1$ and $\omega_2(t)$ are captured by our method, despite the additional, {\em unknown} ``observation'' function.
In addition, the results imply that, indeed, our method empirically recovers an accurate solution even in this ``more complex'' scenario with time-varying spring constant. Moreover, they suggest that the empirical solution obtained by our method may be reliable even in cases for which we have no known ground truth, such as the case described in the main text.
In light of these findings, we believe that our approach can serve as an important, viable, empirical tool in the investigation of nonlinear dynamical systems with time-varying parameters.

\begin{algorithm}[h!]
\caption{Iterative analysis of data arising from dynamical systems}
\textbf{Input:} 3D data tensor $\mathbf{Y}$, initial metrics on samples $\boldsymbol{y}_{v}$ from the variables axis and on samples $\boldsymbol{y}_{t}$ from the time axis , i.e., $\| \cdot \| _{\mathcal{V}}^{(0)}$ and $\| \cdot \| _{\mathcal{T}}^{(0)}$.

\textbf{Construction:}

Set n=0\\

Repeat:

\begin{enumerate}
\item Call Algorithm \ref{algo2} with the metrics $\| \cdot \| _{\mathcal{V}}^{(n)}$ and $\| \cdot \| _{\mathcal{T}}^{(n)}$ as inputs,
and obtain the informed metric $\|\boldsymbol{y}_p-\boldsymbol{y}_l\|_{\mathcal{P}}^{(n+1)}$.
\item Apply diffusion maps with the metric $\|\boldsymbol{y}_p-\boldsymbol{y}_l\|_{\mathcal{P}}^{(n+1)}$ and obtain the $n$-th iteration embedding of samples $\boldsymbol{y}_{p}$ from the parameters axis.

\item Call Algorithm \ref{algo2} with the metrics $\| \cdot \| _{\mathcal{P}}^{(n+1)}$ and $\| \cdot \| _{\mathcal{T}}^{(n)}$ as inputs and obtain the informed metric $\|\boldsymbol{y}_v-\boldsymbol{y}_{\nu}\|_{\mathcal{V}}^{(n+1)}$.

\item Apply diffusion maps with the metric $\|\boldsymbol{y}_v-\boldsymbol{y}_{\nu}\|_{\mathcal{V}}^{(n+1)}$ and obtain the $n$-th iteration embedding of the samples $\boldsymbol{y}_v$ from the variables axis.

\item Call Algorithm \ref{algo2} with the metrics $\|\cdot\|_{\mathcal{P}}^{(n+1)}$ and $\| \cdot \| _{\mathcal{V}}^{(n+1)}$ as inputs and obtain the informed metric $\|\boldsymbol{y}_{t}-\boldsymbol{y}_{\tau}\|_{\mathcal{T}}^{(n+1)}$.

\item Apply diffusion maps with the metric $\|\boldsymbol{y}_{t}-\boldsymbol{y}_{\tau}\|_{\mathcal{T}}^{(n+1)}$ and obtain the $n$-th iteration embedding of samples $\boldsymbol{y}_{t}$ from the time axis.

\item Set $n=n+1$ and jump to Step 1.
\end{enumerate}

\label{algo1}
\end{algorithm}

\begin{algorithm}[h!]
\caption{Informed metric computation based on trees}
\textbf{Input:} 3D data tensor $\mathbf{Y}$, metrics on two axes
$\| \cdot \|_{(a)}$ and $\|\cdot\|_{(b)}$.

\textbf{Output:} A metric on the third axis $\| \cdot \| _{(c)}$.

\textbf{Initialization:}
\begin{enumerate}

\item Construct a partition tree $\mathcal{T}_{(a)}$ of the samples $\boldsymbol{y}_{a}$ based on the metric $\| \cdot \|_{(a)}$.

\item Construct a partition tree $\mathcal{T}_{(b)}$ of the samples $\boldsymbol{y}_{b}$ based on the metric $\| \cdot \|_{(b)}$.

\end{enumerate}

\textbf{Construction:}

\begin{enumerate}

\item Build the filters $g_{I\times J}$ for all $I\in\mathcal{T}_{(a)}$
and $J\in\mathcal{T}_{(b)}$ as in \eqref{eq:filter2d}.

\item Transform the samples $\boldsymbol{y}_{c}$ by computing $\mathcal{F}_{(c)}(\boldsymbol{y}_{c})$
as in \eqref{eq:transform}.

\item Compute the informed metric
\[
\|\boldsymbol{y}_{c}-\boldsymbol{y}_{c'}\|_{(c)}=\|\boldsymbol{y}_{c}-\boldsymbol{y}_{c'}\|_{1}+\gamma\| \mathcal{F}_{(c)}(\boldsymbol{y}_{c})-\mathcal{F}_{(c)}(\boldsymbol{y}_{c'})\|_{1}
\]
between all possible pairs of samples.\end{enumerate}
\label{algo2}
\end{algorithm}

\bibliography{pnas-sample}


\acrodef{EMD}{earth mover's distance}
\acrodef{ODE}{ordinary differential equation}

\end{document}